\documentclass[aps,twocolumn,showpacs,superscriptaddress,preprintnumbers]{revtex4-2}
\usepackage[utf8]{inputenc}
\usepackage{amsmath,amsfonts,amssymb}
\usepackage{graphicx}
\usepackage{float}
\usepackage{subfig}
\usepackage{xcolor}
\usepackage{xspace}

\usepackage{lipsum}

\usepackage{amsmath}
\usepackage{graphicx}
\usepackage[colorinlistoftodos]{todonotes}
\usepackage[colorlinks=true, allcolors=blue]{hyperref}
\usepackage{hyperref}

\hypersetup{
    colorlinks=true,
    linkcolor=blue,
    filecolor=magenta,      
    urlcolor=cyan,
    pdftitle={Sharelatex Example},
    bookmarks=true,
    pdfpagemode=FullScreen,
    }

\newcommand{\unit}[1]{\ensuremath{\mathrm{\,#1}}\xspace}
\newcommand{\e}{\unit{e^{-}}}

\newcommand{\fer}[1]{\textcolor{black}{#1}}

\begin{document}

\title{Fast Single-Quantum Measurement with a \\
Multi-Amplifier Sensing Charge-Coupled Device}

\author{Ana M.\ Botti}
\affiliation{Fermi National Accelerator Laboratory, P.O.\ Box 500, Batavia, IL 60510, USA}
\affiliation{Kavli Institute for Cosmological Physics, University of Chicago, Chicago, IL 60637, USA}

\author{Brenda A. Cervantes-Vergara}
\affiliation{Fermi National Accelerator Laboratory, P.O.\ Box 500, Batavia, IL 60510, USA}
\affiliation{Universidad Nacional Autónoma de México, Ciudad de México, México}

\author{Claudio R.\ Chavez}
\affiliation{Fermi National Accelerator Laboratory, P.O.\ Box 500, Batavia, IL 60510, USA}
\affiliation{Universidad Nacional del Sur (UNS), Bahía Blanca, Argentina}

\author{Fernando Chierchie}
\affiliation{Instituto de Investigaciones en Ingeniería Eléctrica “Alfredo C. Desages” CONICET, Bahía Blanca, Argentina}
\affiliation{Universidad Nacional del Sur (UNS), Bahía Blanca, Argentina}

\author{Alex Drlica-Wagner}
\affiliation{Fermi National Accelerator Laboratory, P.O.\ Box 500, Batavia, IL 60510, USA}
\affiliation{Kavli Institute for Cosmological Physics, University of Chicago, Chicago, IL 60637, USA}
\affiliation{Department of Astronomy and Astrophysics, University of Chicago, Chicago, IL 60637, USA}

\author{Juan Estrada}
\affiliation{Fermi National Accelerator Laboratory, P.O.\ Box 500, Batavia, IL 60510, USA}

\author{Guillermo Fernandez Moroni}
\affiliation{Fermi National Accelerator Laboratory, P.O.\ Box 500, Batavia, IL 60510, USA}

\author{Stephen E. Holland}
\affiliation{Lawrence Berkeley National Laboratory, One Cyclotron Rd, Berkeley, CA 94720, USA}

\author{Blas J.\ Irigoyen Gimenez}
% \email{birigoyen@fiuna.edu.py}
\affiliation{Facultad de Ingeniería, Universidad Nacional de Asunción, San Lorenzo, Paraguay}

\author{Agustin J.\ Lapi}
\affiliation{Instituto de Investigaciones en Ingeniería Eléctrica “Alfredo C. Desages” CONICET, Bahía Blanca, Argentina}
\affiliation{Universidad Nacional del Sur (UNS), Bahía Blanca, Argentina}
\affiliation{Fermi National Accelerator Laboratory, P.O.\ Box 500, Batavia, IL 60510, USA}

\author{Edgar Marrufo Villalpando}
\affiliation{Kavli Institute for Cosmological Physics, University of Chicago, Chicago, IL 60637, USA}
\affiliation{Department of Physics, University of Chicago, Chicago, IL 60637, USA}

\author{Miguel Sofo Haro}
\affiliation{Universidad Nacional de Córdoba, Instituto de Física Enrique Gaviola (CONICET) and Reactor Nuclear RA0 (CNEA), Córdoba, Argentina}

\author{Javier Tiffenberg}
\affiliation{Fermi National Accelerator Laboratory, P.O.\ Box 500, Batavia, IL 60510, USA}

\author{Sho Uemura}
\affiliation{Fermi National Accelerator Laboratory, P.O.\ Box 500, Batavia, IL 60510, USA}

\begin{abstract}
A novel readout architecture that uses multiple non-destructive floating-gate amplifiers to achieve sub-electron readout noise in a thick, fully-depleted silicon detector is presented.
This Multi-Amplifier Sensing Charge-Coupled Device (MAS-CCD) can perform multiple independent charge measurements with each amplifier; measurements with multiple amplifiers can then be combined to further reduce the readout noise.
The readout speed of this detector scales roughly linearly with the number of amplifiers without requiring segmentation of the active area.
The performance of this detector is demonstrated, emphasizing the ability to resolve individual quanta and the ability to combine measurements across amplifiers to reduce readout noise.
The unprecedented low noise and fast readout of the MAS-CCD make it a unique technology for astronomical observations, quantum imaging, and low-energy interacting particles.

\end{abstract}

\maketitle
\section{Introduction}

Silicon semiconductor sensors with sub-photon or sub-electron resolution are a major scientific breakthrough \cite{Simoen_1999, janesick_1990,boukhayma2017ultra}. 

While some technologies utilize signal amplification by charge multiplication \cite{Hynecek2003, BUZHAN2003}, others have demonstrated single-photon detection using small sensing structures with high charge-to-voltage gain \cite{Fossum_2016}.
A third class of detectors performs nondestructive readout to reduce the readout noise by averaging several measurements of the same collected charge. The non-destructive readout was originally enabled by a floating-gate amplifier (FGA)~\cite{Wen_1973, Wen_1974, FG_1975} and later proposed as the readout stage of a novel Charge Coupled Device (CCD), named the Skipper CCD \cite{janesick_1990, Janesick_patent, Chandler_1990, Hynecek_1997}. Recent results from $p$-channel, fully-depleted Skipper CCD sensors fabricated with high-resistivity silicon \cite{Holland:2003, HV_2006, HV_2009} demonstrate that the readout noise can be reduced to arbitrarily low levels \cite{skipper_2012, Tiffenberg:2017aac, cancelo2021low} with extremely low dark current \cite{cababie_2022, Barak2020} and high quantum efficiency for blue to near-infrared photons \cite{Drlica_2020}. 

The main drawback of the Skipper CCDs is their slow readout speed at the desired noise level.  In recent years, many applications such as dark matter search \cite{Barak2020, OSCURA2020}, neutrino detection \cite{violeta2020}, and studies of fundamental properties of silicon \cite{Rodrigues_2020, botti_2022} have exploited the capability of Skipper CCDs despite this caveat; however, other applications such as quantum imaging \cite{estrada2021ghost}, astronomical instrumentation \cite{Drlica_2020, RauscherNASA2019}, and sub-shot-noise microscopy \cite{Samantaray2017}, cannot fully profit from the Skipper CCD due to the long readout time. This work presents an experimental demonstration of a new readout architecture that leverages multiple non-destructive readout stages to reduce readout noise and increase readout speed. The technology is called Multi-Amplifier Sensing Charge Coupled Device (MAS-CCD) \cite{holland_2023}.

\section{Detector Architecture}
\label{sec:architecture}

\begin{figure*}[t]
    \centering
    \includegraphics[width=1\textwidth]{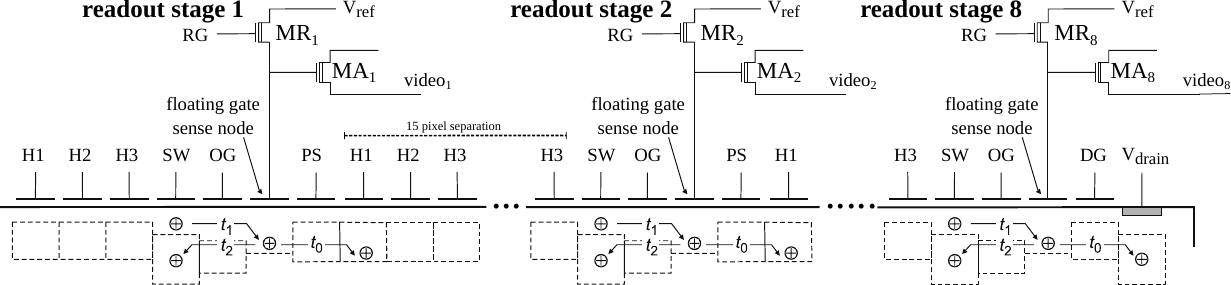}
    \caption{Architecture of the eight inline amplifiers at the end of the serial register of the MAS-CCD.}
    \label{fig:architecture}
\end{figure*}

The MAS-CCD uses a series of output amplifiers capacitively connected to the sensor channel (Fig.~\ref{fig:architecture}) via a floating gate (FG), which allows for non-destructive charge measurements as in the Skipper CCDs \cite{Tiffenberg:2017aac}. Furthermore, the MAS setup also enables charge transport through this output register, typically called the serial register, without degradation, so the pixel charge can be measured multiple times by multiple amplifiers (MA$_1$, $\dots$, MA$_8$ in Fig.~\ref{fig:architecture}) in each output stage. The dump gate (DG) and drain (V$_{\text{drain}}$) contacts in the last stage remove the charge from the channel of the serial register after its measurement in the last amplifier. The intermediate amplifiers use the pixel-separation gates (PS) to facilitate the charge removal from the sense node and transfer to the next amplifier. The H1, H2, and H3 gates provide the three-clock sequence to move the charge between consecutive stages. During normal operation, each amplifier simultaneously measures the charge packets from different pixels. The final pixel value is calculated by averaging the available samples from each of the amplifiers

\begin{equation}
    \text{pixel value} = \frac{1}{N_{a}}\frac{1}{N_s}\sum^{N_{a}}_{j=1}\sum^{N_{s}}_{i=1}s_{j,i},
\end{equation}

\noindent where $s_{j,i}$ is the charge measurement sample $i$ from the amplifier $j$, $N_{a}$ is the number of amplifiers in the serial register, and $N_s$ is the number of samples taken with each amplifier. Assuming that the readout noise of each amplifier is independent and similar in standard deviation ($\sigma_0$), the standard deviation of the readout noise in the final measurement is,

\begin{equation}
    \sigma = \frac{\sigma_0}{\sqrt{N_s}\sqrt{N_{a}}},
\end{equation}
which has an additional reduction factor ($\sqrt{N_a}$) compared to the Skipper-CCD \cite{Tiffenberg:2017aac}.

In the fastest readout mode, each amplifier measures the charge once ($N_s=1$), and the noise reduction arises from the combination of measurements from different amplifiers.

 \cite{holland_2023} taking into account the architecture presented in Fig.~\ref{fig:architecture}, is
Following \cite{holland_2023}, the MAS-CCD architecture presented in Fig.~\ref{fig:architecture} with $N_{a}$ inline amplifiers each separated by $k_{\text{iAmp}}$ and an extended pixel region of $k_{\text{ex}}$ will read the first pixel in the serial register with all amplifiers after a time

\begin{equation}
t^{\text{MAS}}_{1} = k_{\text{ex}}t_{\text{shift}} + t_{\text{shift}} +  t_{\text{read}} + (N_{a}-1)k_{\text{iAmp}}(t_{\text{shift}} + t_{\text{read}}),
\end{equation}

\noindent where $t_{\text{shift}}$ is the time needed to shift the charge by one serial pixel and $t_{\text{read}}$ is the time to perform one non-destructive read.
Assuming an active region of $N_\text{col}$, it will take ($N_\text{col}-1$) shifts and pixel readouts to read the remaining pixels from the first row of the CCD. 
Thus, the total time to read the row is

\begin{equation}
    t^{\text{MAS}}_{\text{row}} = t^{\text{MAS}}_{1} + (N_\text{col}-1)(t_{\text{shift}} + t_{\text{read}}).
\end{equation}
The time to read an array of $N_\text{row}$ rows is $N_\text{row}$ multiplied by the sum of $t_\text{row}$ and the row shift time.

For a conventional CCD with one amplifier in the serial register taking one measurement per pixel, the readout time for one row is $t^{\text{CCD}}_{\text{row}} = k_{\text{ex}}t_{\text{shift}} + N_\text{col}(t_{\text{shift}} + t_{\text{read}})$. The ratio of time increase is 
~
\begin{equation}
(t^{\text{MAS}}_{\text{row}} - t^{\text{CCD}}_{\text{row}})/t^{\text{CCD}}_{\text{row}} \approx    (N_{a}-1)k_{\text{iAmp}}/N_\text{col},
\end{equation}
~
where it is assumed that the number of extended pixels is much smaller than the total number of columns of the array, $k_{\text{ex}} \ll N_{col}$.

When the number of pixels in the serial amplifier chain, $(N_{a}-1)k_{\text{iAmp}}$, is much smaller than $N_\text{col}$, the readout time is similar to that of a single output sensor, but with a noise reduction factor of $\sqrt{N_a}$. 
For example, for a MAS-CCD with a square active region of 4 million pixels, the extra readout time needed for eight amplifiers (sixteen amplifiers) separated by $ k_{\text{iAmp}} = 15$ pixels is approximately 5\% (11\%) for a noise reduction of $\sim 2.8\times$ ($4\times$). A faster readout scheme can be used if, during the readout operation, a row of the active region is dumped into the serial register just after the $N_\text{col}$ pixels in the serial register are empty event if the charge from the last pixels of the previous row are being read in the pixels of the output stages.

\begin{figure}[t]
    \centering
    \includegraphics[width=0.48\textwidth]{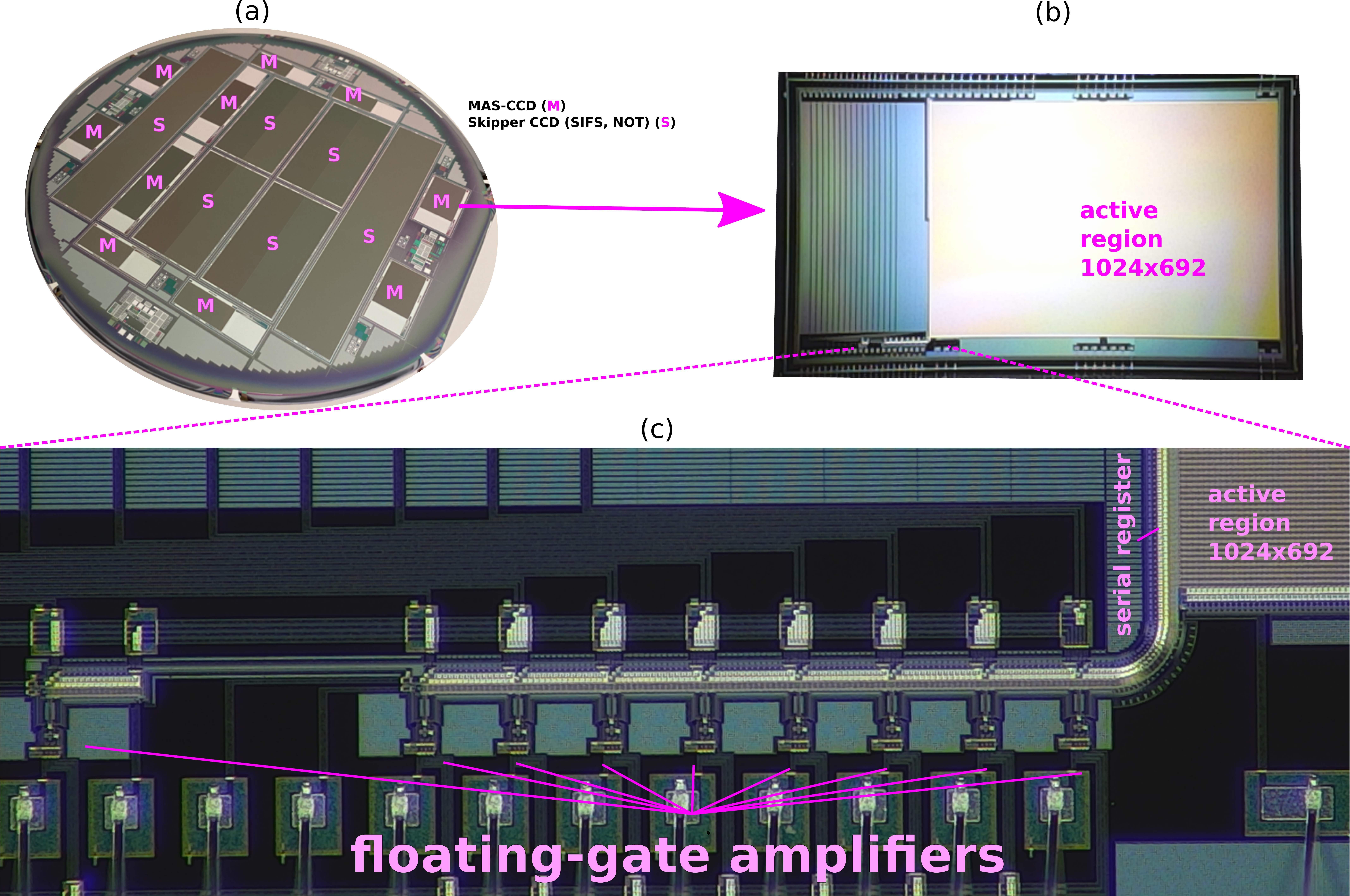}
    \caption{A picture of a 150 mm diameter and 675 $\mu$m thick silicon wafer containing several MAS-CCD and Skipper CCD architectures. For more details about the fabricated sensors in this wafer see \cite{holland_2023}. (b) Image of the eight-amplifier MAS-CCD under study for this work. (c) Microscope image of the eight floating gate amplifiers at the end of the serial register together with an additional amplifier for differential output.}
    \label{fig:microscope_view}
\end{figure}

Under the auspices of the DOE Quantum Information Science initiative, $p$-channel 675$\mu$m-thick MAS-CCDs were fabricated on high-resistivity $n$-type silicon (${\sim}\,10$ k$\Omega$\,cm) with eight and sixteen amplifiers (Fig.~\ref{fig:microscope_view}a). The sensors were designed at LBNL to be operated as thick fully-depleted devices with high quantum efficiency over a broad wavelength range \cite{Holland:2003} (see \cite{holland_2023} for more details on the fabricated sensors). The sensors were fabricated at Teledyne DALSA Semiconductor, diced at LBNL, and packaged/tested at Fermilab. In addition, some of the wafers were fabricated following a hybrid fabrication production model used for DECam and DESI CCDs \cite{HOLLAND2007653}, where specialized processing of the back surface and metalization was done at LBNL MicroSystems Laboratory to produce 250$\mu$m thick back-illuminated wafers. The results presented in this article come from thick sensors that do not have that specialized back treatment.

The eight-amplifier MAS-CCD used for the results presented in this article is shown in Fig.~\ref{fig:microscope_view}b. 
The sensor consists of a 1024 $\times$ 692 array of 15\,$\mu$m $\times$ 15\,$\mu$m pixels and a substrate thickness of 675\,$\mu$m.
The $N_a = 8$ output amplifiers are separated by $k_{\text{iAmp}} = 15$ pixels with a prescan of $k_{\text{ex}} = 27$ pixels (Fig.~\ref{fig:microscope_view}c). The pixels in the prescan are used to bend the serial register by 90 degrees as shown in the picture.
The detector has an additional floating gate amplifier disconnected from the serial register for differential output operation, and noise and clock feedthrough sensing. A floating diffusion output stage \cite{janesick2001scientific} is situated in the opposite corner of the sensor but was not used in this work.

\section{Experimental results}
\label{sec:measurements}

The operational demonstration of the MAS-CCD focuses on the two distinct features with respect to the single FG output stage of a Skipper CCD: (1) the lossless transfer of charge between independent amplifiers, and (2) the reduction of noise by combining measurements from different amplifiers. The low-threshold acquisition (LTA) controller \cite{cancelo2021low} was used to read the device. Initially, a single LTA was used to instrument three channels to study the charge transfer properties between stages. Subsequently, two synchronized LTAs were utilized to instrument all eight channels and perform  noise reduction tests. The sensor was operated in a fully depleted mode using a substrate voltage of 70\,V. Measurements were collected at an operating temperature of 140\,K.

\begin{figure}[t]
    \centering
    \includegraphics[width=0.8\columnwidth, trim={3cm 1cm 3cm 1cm}, clip]{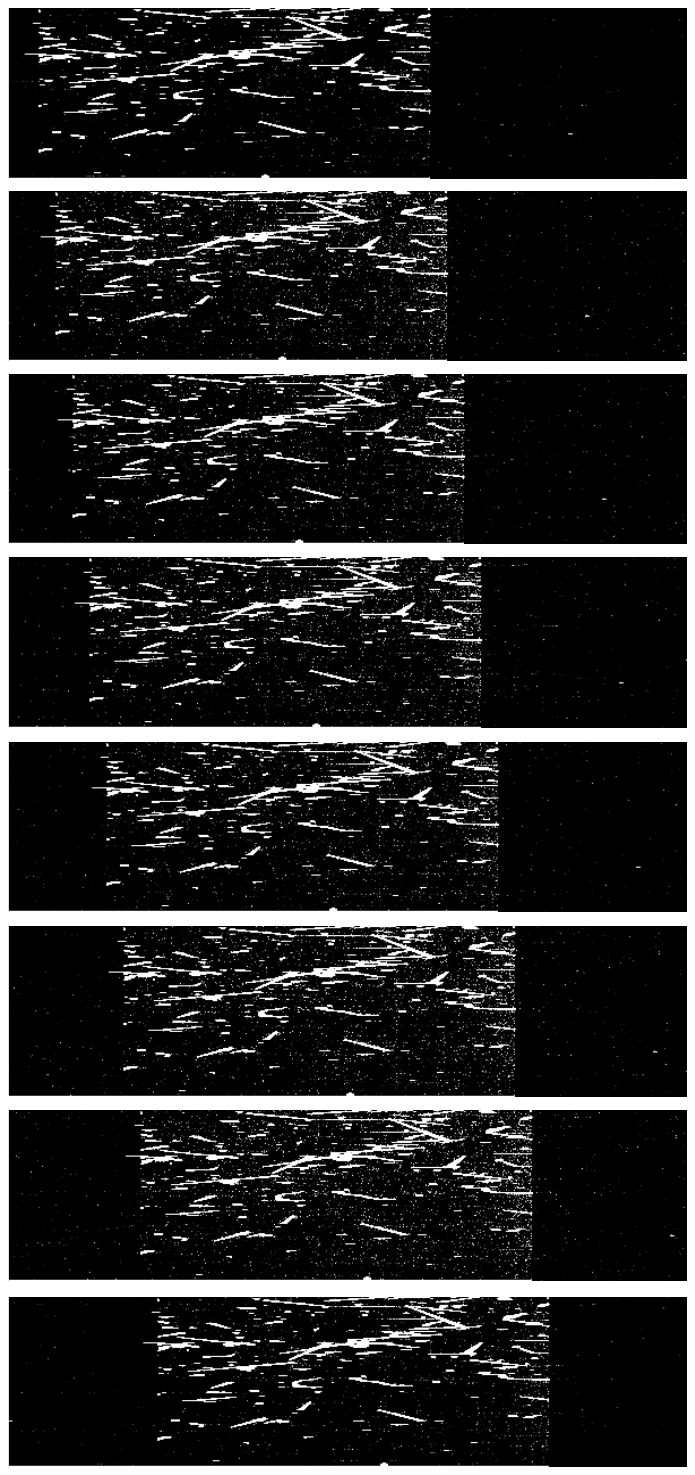}
    \vspace{-1em}
    \caption{Image obtained by each of the eight MAS-CCD amplifiers. The active region in the center of the image is populated by muon tracks (white streaks). The dark region to the left represents the distance in pixels of the serial register between each stage and the active region ($jk_{\text{iAmp}} + k_{\text{ex}}$ for the output stage $j$), while the dark region of the right is the overscan pixels with arbitrary width for this test. The output images are sorted from top to bottom for the first to eight amplifiers. The images were taken using a vertical binning \cite{janesick2001scientific} of ten pixels. }
    \label{fig:all_channels}
\end{figure}

Figure~\ref{fig:all_channels} displays an example of the images obtained with the eight MAS-CCD amplifiers. The active region is the rectangular portion located in the center of the image, which features muon tracks. The black pixels on the left of this region represent the prescan of each amplifier; the width of this region increases for amplifiers located farther from the active area. The black region on the right is the virtual overscan for each amplifier. Notably, the same muon tracks are observed with each amplifier.

\subsection{Charge transfer in the serial register}

\begin{figure}[t]
\centering
\includegraphics[width=0.48\columnwidth]{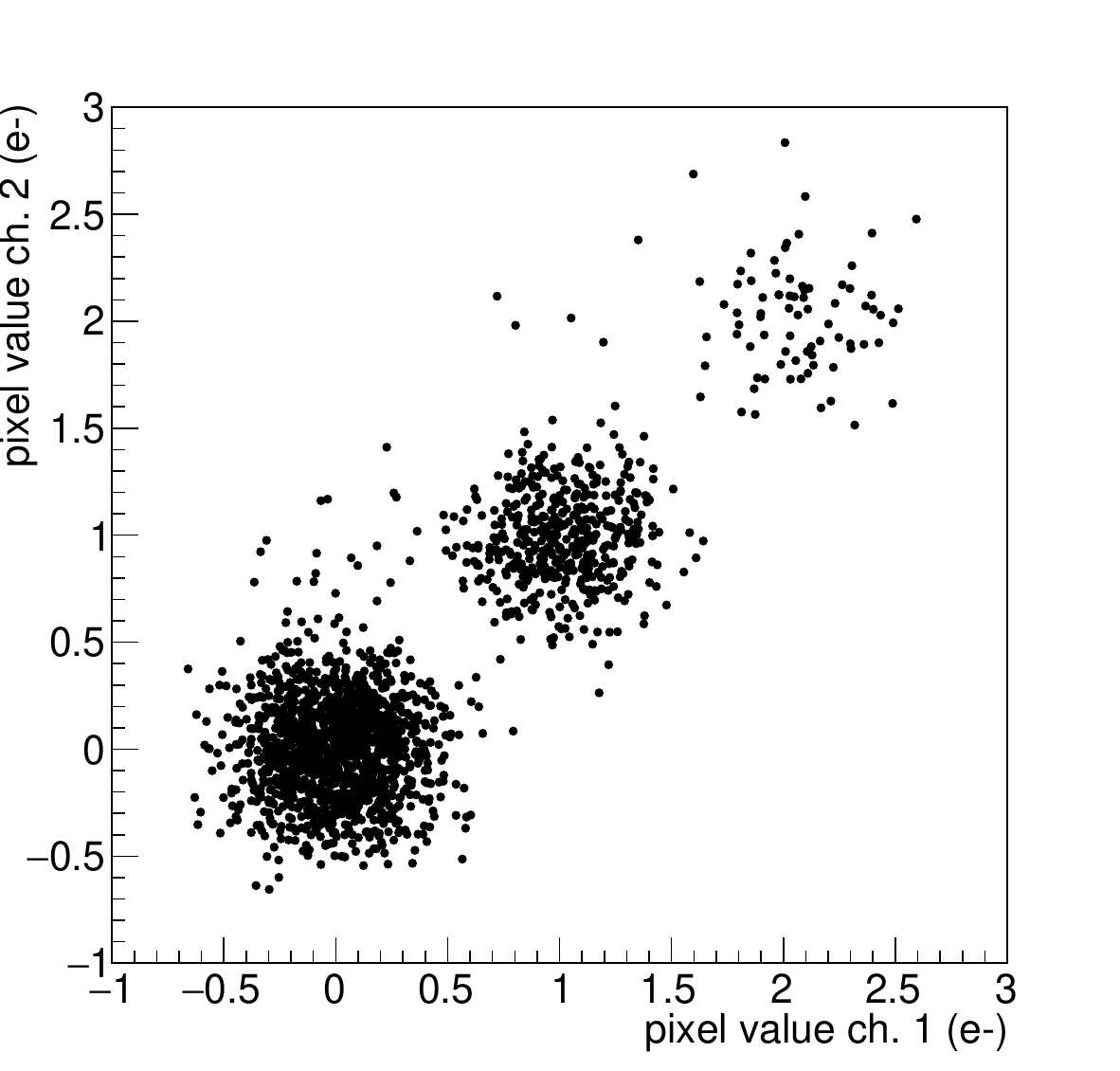}
\includegraphics[width=0.48\columnwidth]{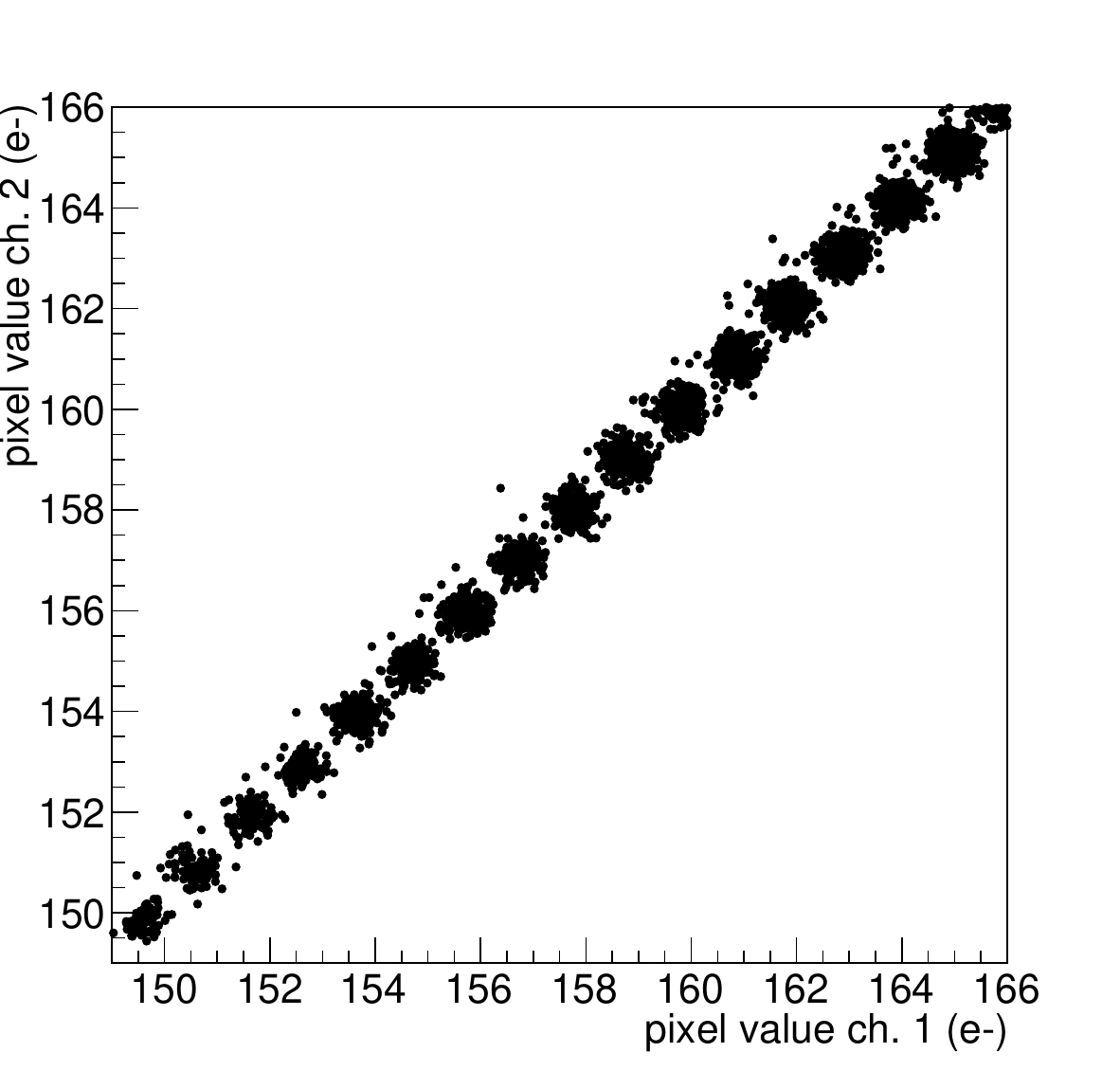}

    \caption{Scatter plot of the measured charge per pixel (in units of e$^-$) for two different channels of the MAS-CCD output stage in the low (left) and moderate (right) charge regimes. The quantized clustering of points demonstrates consistent single-electron resolution by both amplifiers. The lack of outliers shifting down demonstrates high charge transfer efficiency between consecutive stages.}
    \label{fig:charge_comparison}
\end{figure}

Studies of the charge transfer efficiency are presented in Fig.~\ref{fig:charge_comparison}. The scatter plots show the charge measured in each pixel by the first and second amplifiers in the readout chain for two ranges: between 1--3 electrons (left) and 150--166 electrons (right). The sensor is operated in the sub-electron noise regime ($\sigma \sim 0.08$ e$^{-}$ rms/pix; $N_s$ = 1024) such that individual charge carriers are quantized. The clouds of points represent the number of charge carriers measured by each amplifier, and the width of these clusters corresponds to the Gaussian readout noise of the measurement. The plots show a good match between measurements in both amplifiers and ranges. Charge loss between the first and second amplifiers would appear as points shifted down by one (or more) electrons from the one-to-one distribution. On the other hand, some measurements show a charge increase of one electron between the first and second amplifiers, which can be explained by thermal radiation from the room-temperature wall of the vacuum chamber hitting the serial register. Since each amplifier performs $N_s = 1024$ samples, there is ample time for radiation to collect in the serial register and produce this extra charge. 

\subsection{Noise reduction}

The noise reduction capability of the MAS-CCD is demonstrated by combining measurements from different amplifiers. 

\begin{figure}[t]
    \centering
    
    \includegraphics[width=0.90\columnwidth]{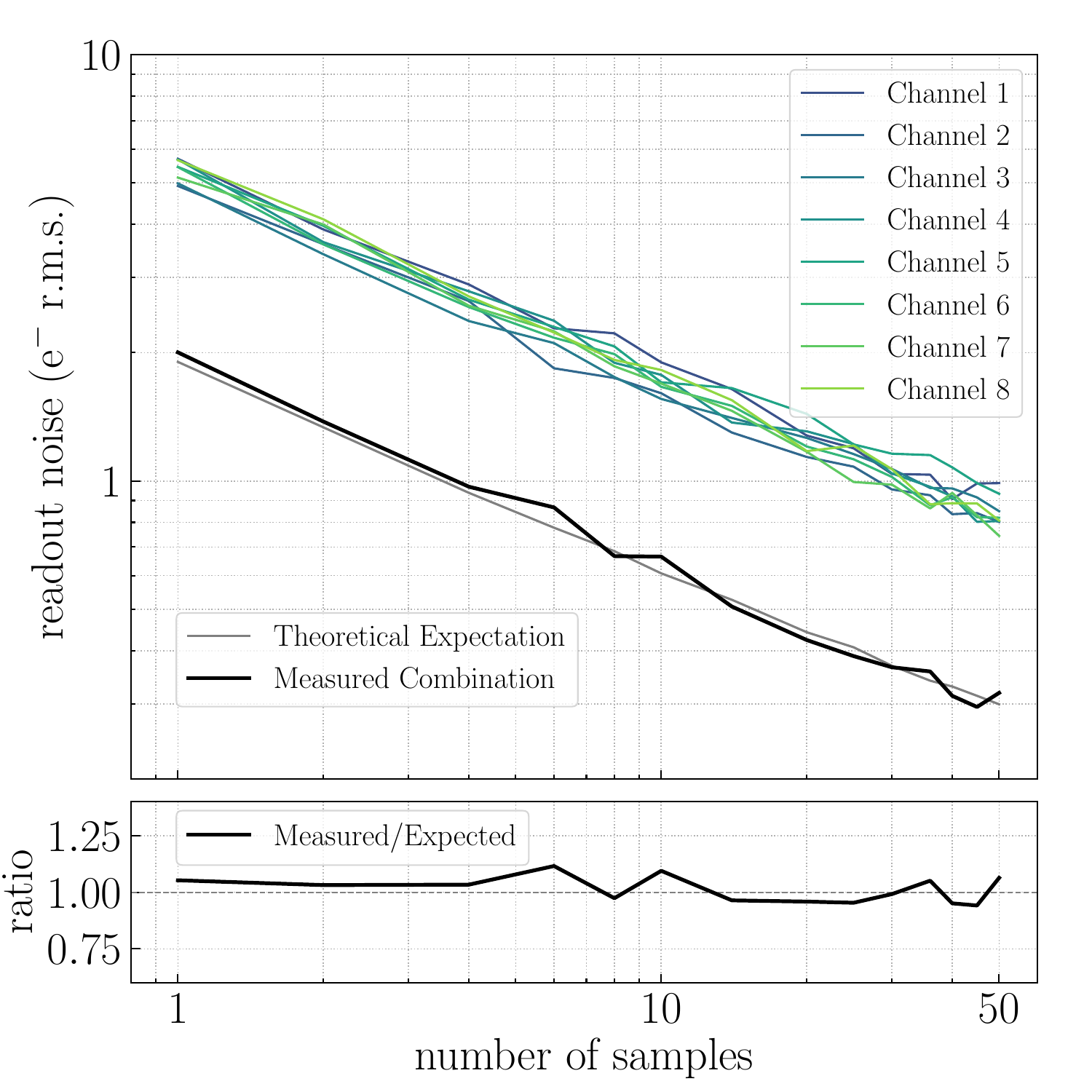}
    \caption{Top: Standard deviation of the readout noise as a function of the number of measurements taken by each amplifier (green-scale lines). The gray line shows the expected noise reduction from combining all channels assuming the noise contributions are independent. The black line shows the actual combination. The numbers of samples per amplifier used to measure the noise are 1, 2, 4, 6, 8, 10, 14, 20, 25, 30, 36, 40, 45, 50.
    Bottom: Ratio between the measured standard deviation after combining the information from all channels and the expectation assuming the noise contributions from the channels are independent.} 
    
    \label{fig:standard_dev_reduction}
\end{figure}

Fig.~\ref{fig:standard_dev_reduction} shows the measured readout noise (standard deviation measured for overscan pixels) for each channel as a function of the number of measurements performed by each channel. This test is focused on a small number of samples taken for each amplifier targeting astronomy applications where long readout times are not desirable. All amplifiers exhibit similar single-sample noise ($\sigma_0$), and the individual noise is reduced by increasing the number of samples.
The single-sample noise ranges from $4.9 \lesssim \sigma_0 \lesssim 5.7$ \e\,rms for an integration time in the pedestal and signal levels of 23.3 $\mu$s, which is slightly higher than previous measurements on the Skipper CCDs \citep{cancelo2021low}. This is due to the fact that existing electronics were used in the signal chain and were not optimized for noise reduction. The long integration time was explored to evaluate the noise reduction by combining measurements from different amplifiers when the contribution of the white noise is highly suppressed \cite{janesick2001scientific}. Since the samples of one pixel are taken at different times at the different amplifiers, correlated noise in the readout chain might be a source of loss of noise reduction power when combining the samples. 

Next, measurements of the noise reduction after combining the results from different amplifiers are presented. The black curve in Fig.~\ref{fig:standard_dev_reduction} represents the measured standard deviation after averaging the pixel information as a function of the number of samples in each amplifier. Although the readout chain is not fully optimized, the final readout noise is smaller than any of our previous results with Skipper-CCD. The grey curve shows the theoretical expectation after combining the individual measurements assuming that the noise is uncorrelated for the different amplifiers. In this case the noise should be reduced after averaging to:
\begin{equation}
\label{eq: expected noise}
    \sigma =\dfrac{\sqrt{\sum_{j=1}^{N_a}(\sigma_{j}^2)}}{N_a}.
\end{equation} 
where $\sigma_{j}$ is the noise standard deviation of amplifier $j$ after averaging all its available samples. The lower panel of Fig.~\ref{fig:standard_dev_reduction} shows the ratio between the standard deviation of the combined measurement and the expected one from Eq.~\ref{eq: expected noise}. The measured noise is in good agreement with the theoretical expectation.

\begin{figure}[th]
    \centering
    \includegraphics[width=\columnwidth, trim={0cm 0cm 0cm 1cm}, clip]{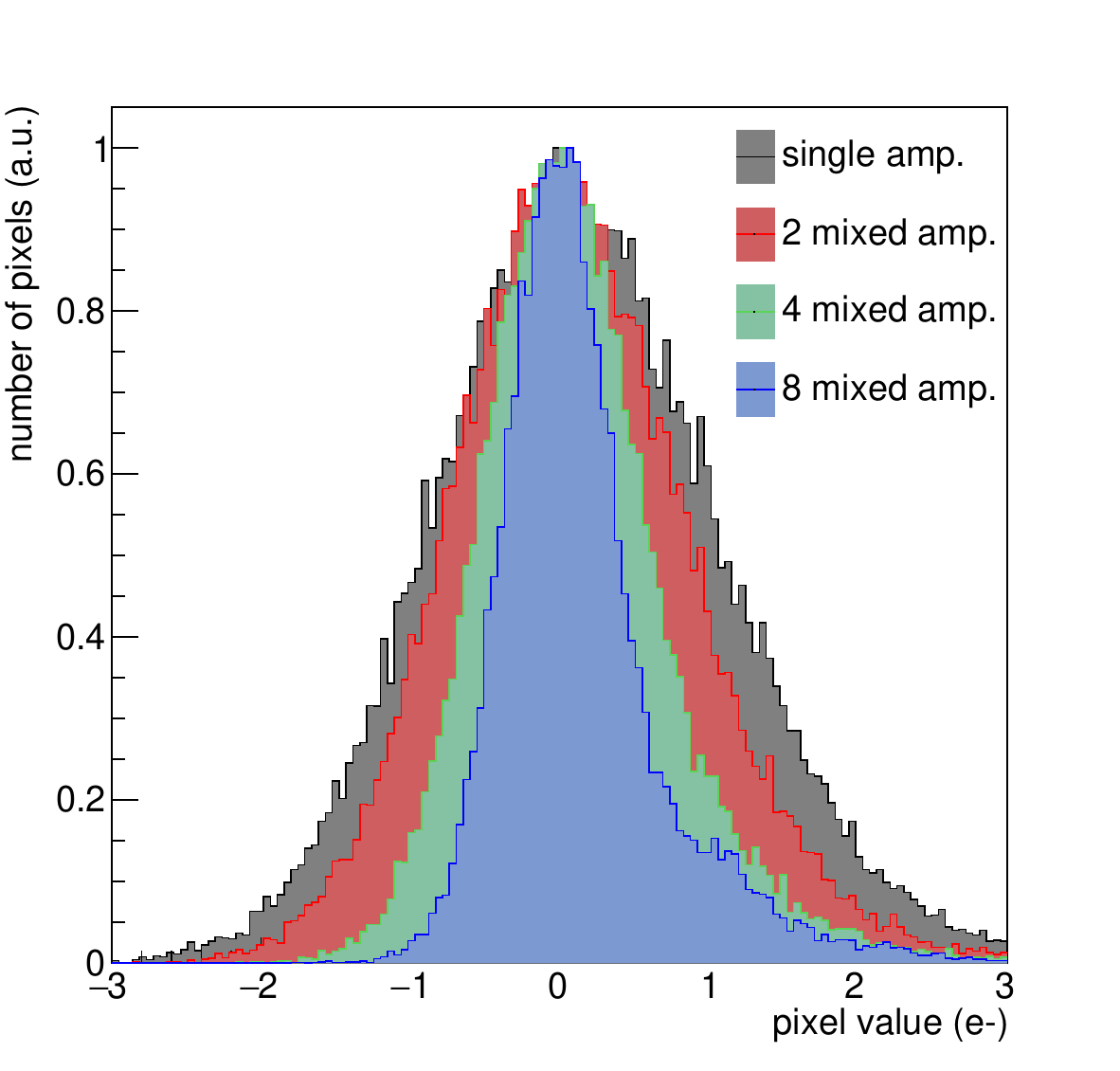}
    \caption{Histogram of the values of a set of overscan pixels using the information by one, two, four, and eight independent readout amplifiers. Each amplifier is performing 50 independent measurements of the charge. When measurements from all eight amplifiers are combined, evidence for charge quantization can be seen through the emergence of a bump corresponding to pixels containing one electron.
    \label{fig:histogram_noise_reduction}
    }
\end{figure}

The noise reduction is illustrated in Fig.~\ref{fig:histogram_noise_reduction}. The four charge histograms come from the same pixels, but use different numbers of channels to calculate the pixel value performing $N_s=50$ measurements in each amplifier. As the number of amplifiers increases, the standard deviation of the pixel noise decreases. When all eight amplifiers are combined (blue histogram), the noise is reduced to reveal charge quantization (a bump for pixels containing one electron). 

\begin{figure}[h]
    \centering
    \includegraphics[width=\columnwidth]{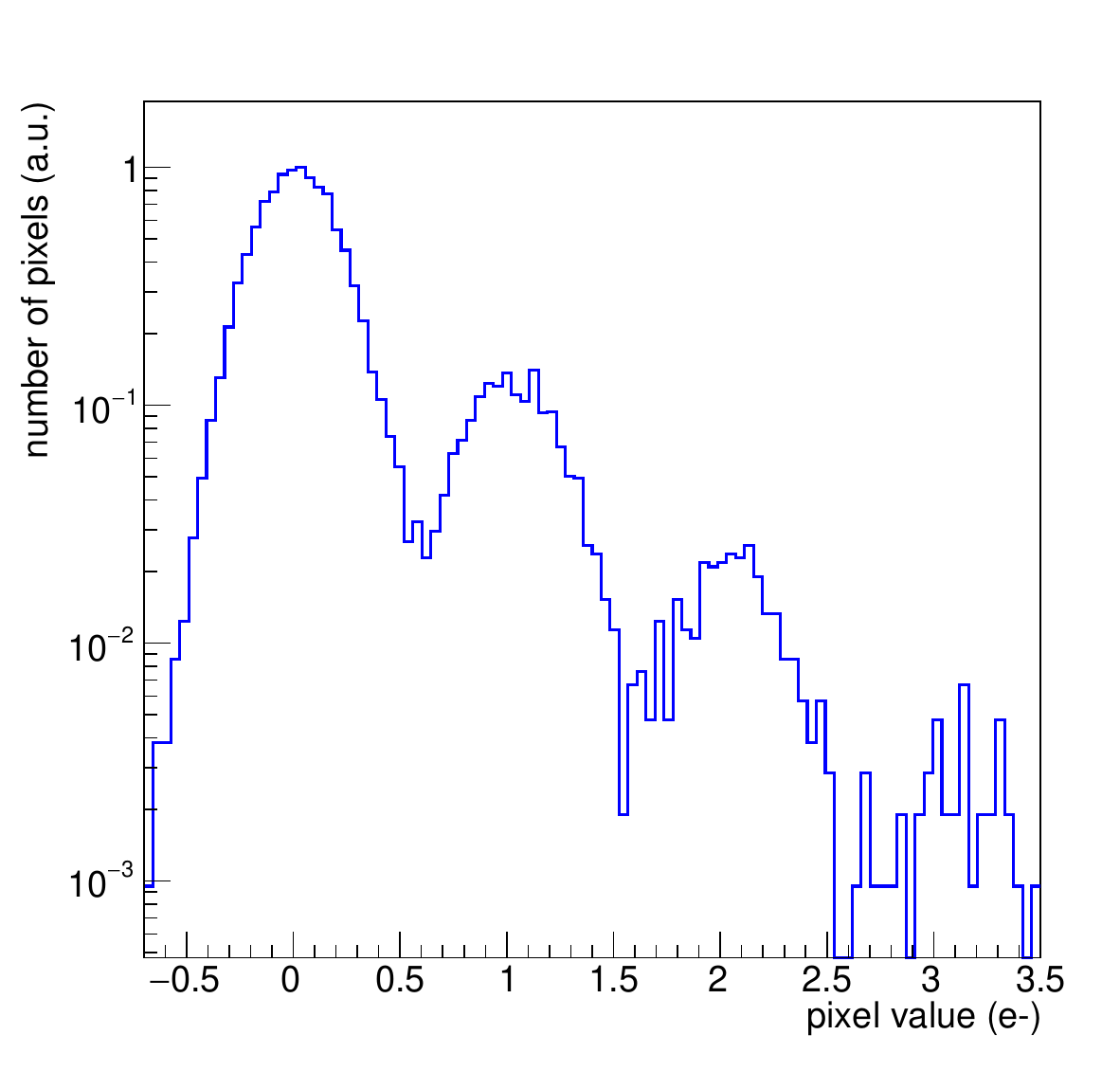}
    \caption{Single-carrier resolution demonstrated by combining measurements from eight serial amplifiers, each performing 400 non-destructive measurements of the charge in each pixel.}
    \label{fig:histogram_nsamp1000}
\end{figure}

In a more extreme case when 400 samples are taken per amplifier, the measurements combined across all amplifiers in Fig.~\ref{fig:histogram_nsamp1000}, show a clear distribution of peaks centered in the number of carriers per pixel. The distance between peaks is used to calculate the gain of the system for the measurements presented here. The symmetry in the peaks proves the conservation of the charge packets during their time in the readout stage. 

\section{Scientific Applications}

\label{sec:applications}

\subsubsection*{Astronomical Observations}

The potential to achieve sub-electron noise with fast readout time is attractive for many areas of astronomy \cite{Tiffenberg:2017aac,Drlica_2020,Maruffo_2022}. 
Two areas of particular interest are high-cadence imaging, where individual exposures are often photon-limited, and spectroscopy, where photons from faint astronomical sources are dispersed over a large detector area.

In both cases, sub-electron readout noise can provide significant increases in sensitivity; however, long readout times can be prohibitive for such observations \cite{Drlica_2020}. 

One specific astronomical application is the direct spectroscopy of extra-solar planets to detect the signatures of life, a major objective of the astronomical community in the coming decades \cite{Astro2020}.
Such observations require deep sub-electron readout noise and high quantum efficiency for near-infrared photons to search for spectral lines from water vapor in the atmosphere of potentially habitable planets \cite{crill2018exoplanet}.
The radiation tolerance of the $p$-channel detectors described here makes them extremely attractive for space-based applications \citep{Dawson:2008}. 
However, in order to avoid saturating the detector with cosmic-ray tracks in the intense radiation environment of space, sensors must be made to work with readout times of $\leq$20 seconds \cite{Rauscher_internal_2019}.

Wide-field, massively-multiplexed spectroscopy of faint stars and galaxies presents another exciting application for the MAS-CCD \citep{Schlegel:2022}.
Readout noise can contribute a significant fraction of the noise budget for observations of faint sources, particularly at blue wavelengths where the sky background level is low \citep{Drlica_2020}.
Future massively-multiplexed spectroscopic instruments need detectors with low readout noise ($\sim$\,1 e$^-$ rms/pix), large format (4k $\times$ 4k pixels), fast readout time ($\lesssim$ 60s), and high quantum efficiency over a broad wavelength range \citep{Asaadi:2022,Schlegel:2022}.
Such requirements could be satisfied by a large format MAS-CCD performing single-charge measurements by 16 serial amplifiers in 4 readout quadrants.

\subsubsection*{Quantum imaging techniques}

Spatial entanglement has been extensively explored for quantum communication \cite{YUAN20101}. Spontaneous parametric down-conversion crystals have eased the production of entangled pairs over a large number of positions \cite{malygin_1985} \fer{for example for ghost imaging \cite{Padgett_2017,Pittman_1995}. Intensity correlation can be used for several other imaging techniques \cite{defienne_2010} such as fluorescence correlation spectroscopy \cite{Bestvater_2010}. }Two-dimensional semiconductor devices provide a good sensor solution for these applications \cite{Orieux_2017}. In particular, the single-photon counting capability and large quantum efficiency of the MAS-CCD make it a promising technology in the field \cite{estrada2021ghost}.  

\subsection{Low-energy interacting particles}

Current applications using CCDs to detect faint energy depositions by particle interactions can be benefited by a faster readout time of the sensors by allowing for shorter net exposure times of the active region. For light galactic dark matter searches using CCDs \cite{Barak2020, DAMIC2019, damicM_2023, oscura2022} applications, a faster readout mode eliminates the probability of piling up carriers from single-carrier sources mimicking a particle interaction of two or more collected charges. For the detection of neutrino generated in nuclear reactors \cite{connie_2022, skipper_above_ground_2022}, it also provides a way to reduce the combination of single carriers that could mimic neutrino interactions and lower the loss of exposure time due to the occupancy of high-energy tracks.
Background reduction is also desirable for the construction of equipment to monitor very low-energy radiation such as the detection of tritium contaminant in water samples as it was explored as part of the GRAIL project \cite{grail}. In this case, low-energy beta by radioactive decay of the tritium molecules is detected using a CCD.

\section{Conclusions}

Sub-electron readout noise with a novel multi-amplifier architecture silicon detector, dubbed the MAS-CCD, was demonstrated.
This detector uses a series of floating gate amplifiers to perform multiple, non-destructive measurements of each pixel charge.
The ability to non-destructively transfer charge between amplifiers and the combination of measurements from individual amplifiers to reduce the noise was demonstrated.
For large devices with small inter-amplifier separations, the MAS-CCD enables an increase in readout speed that is nearly linear to the number of amplifiers.
Exciting applications for such a detector exist in astronomy, quantum imaging, and low-energy interacting microscopy.

\begin{acknowledgments}

The fully depleted Skipper CCD was developed at Lawrence Berkeley National Laboratory, as were the designs described in this work. The CCD development work was supported in part by the Director, Office of Science, of the U.S. Department of Energy under No. DE-AC02-05CH11231.

This research has been partially supported by Javier Tiffenberg's DOE Early Career research program.
This research has been sponsored by the Laboratory Directed Research and
Development Program of Fermi National Accelerator Laboratory (L2019-011, L2022.053), managed by
Fermi Research Alliance, LLC for the U.S. Department of Energy.

This work was partially supported by NASA APRA award No.\ 80NSSC22K1411.
This manuscript has been authored by Fermi Research Alliance, LLC, under contract No. DE-AC02-07CH11359 with the US Department of Energy, Office of Science, Office of High Energy Physics. The United States Government retains and the publisher, by accepting the article for publication, acknowledges that the United States Government retains a non-exclusive, paid-up, irrevocable, worldwide license to publish or reproduce the published form of this manuscript, or allow others to do so, for United States Government purposes.

\end{acknowledgments}

% References
\bibliography{main.bib} 

\begin{thebibliography}{51}%
\makeatletter
\providecommand \@ifxundefined [1]{%
 \@ifx{#1\undefined}
}%
\providecommand \@ifnum [1]{%
 \ifnum #1\expandafter \@firstoftwo
 \else \expandafter \@secondoftwo
 \fi
}%
\providecommand \@ifx [1]{%
 \ifx #1\expandafter \@firstoftwo
 \else \expandafter \@secondoftwo
 \fi
}%
\providecommand \natexlab [1]{#1}%
\providecommand \enquote  [1]{``#1''}%
\providecommand \bibnamefont  [1]{#1}%
\providecommand \bibfnamefont [1]{#1}%
\providecommand \citenamefont [1]{#1}%
\providecommand \href@noop [0]{\@secondoftwo}%
\providecommand \href [0]{\begingroup \@sanitize@url \@href}%
\providecommand \@href[1]{\@@startlink{#1}\@@href}%
\providecommand \@@href[1]{\endgroup#1\@@endlink}%
\providecommand \@sanitize@url [0]{\catcode `\\12\catcode `\$12\catcode
  `\&12\catcode `\#12\catcode `\^12\catcode `\_12\catcode `\%12\relax}%
\providecommand \@@startlink[1]{}%
\providecommand \@@endlink[0]{}%
\providecommand \url  [0]{\begingroup\@sanitize@url \@url }%
\providecommand \@url [1]{\endgroup\@href {#1}{\urlprefix }}%
\providecommand \urlprefix  [0]{URL }%
\providecommand \Eprint [0]{\href }%
\providecommand \doibase [0]{http://dx.doi.org/}%
\providecommand \selectlanguage [0]{\@gobble}%
\providecommand \bibinfo  [0]{\@secondoftwo}%
\providecommand \bibfield  [0]{\@secondoftwo}%
\providecommand \translation [1]{[#1]}%
\providecommand \BibitemOpen [0]{}%
\providecommand \bibitemStop [0]{}%
\providecommand \bibitemNoStop [0]{.\EOS\space}%
\providecommand \EOS [0]{\spacefactor3000\relax}%
\providecommand \BibitemShut  [1]{\csname bibitem#1\endcsname}%
\let\auto@bib@innerbib\@empty
%</preamble>
\bibitem [{\citenamefont {Simoen}\ and\ \citenamefont
  {Claeys}(1999)}]{Simoen_1999}%
  \BibitemOpen
  \bibfield  {author} {\bibinfo {author} {\bibfnamefont {E.}~\bibnamefont
  {Simoen}}\ and\ \bibinfo {author} {\bibfnamefont {C.}~\bibnamefont
  {Claeys}},\ }\href {\doibase https://doi.org/10.1016/S0038-1101(98)00322-0}
  {\bibfield  {journal} {\bibinfo  {journal} {Solid-State Electronics}\
  }\textbf {\bibinfo {volume} {43}},\ \bibinfo {pages} {865} (\bibinfo {year}
  {1999})}\BibitemShut {NoStop}%
\bibitem [{\citenamefont {Janesick}\ \emph {et~al.}(1990)\citenamefont
  {Janesick}, \citenamefont {Elliott}, \citenamefont {Dingiziam}, \citenamefont
  {Bredthauer}, \citenamefont {Chandler}, \citenamefont {Westphal},\ and\
  \citenamefont {Gunn}}]{janesick_1990}%
  \BibitemOpen
  \bibfield  {author} {\bibinfo {author} {\bibfnamefont {J.~R.}\ \bibnamefont
  {Janesick}}, \bibinfo {author} {\bibfnamefont {T.~S.}\ \bibnamefont
  {Elliott}}, \bibinfo {author} {\bibfnamefont {A.}~\bibnamefont {Dingiziam}},
  \bibinfo {author} {\bibfnamefont {R.~A.}\ \bibnamefont {Bredthauer}},
  \bibinfo {author} {\bibfnamefont {C.~E.}\ \bibnamefont {Chandler}}, \bibinfo
  {author} {\bibfnamefont {J.~A.}\ \bibnamefont {Westphal}}, \ and\ \bibinfo
  {author} {\bibfnamefont {J.~E.}\ \bibnamefont {Gunn}},\ }in\ \href {\doibase
  10.1117/12.19452} {\emph {\bibinfo {booktitle} {Charge-Coupled Devices and
  Solid State Optical Sensors}}},\ Vol.\ \bibinfo {volume} {1242},\ \bibinfo
  {editor} {edited by\ \bibinfo {editor} {\bibfnamefont {M.~M.}\ \bibnamefont
  {Blouke}}},\ \bibinfo {organization} {International Society for Optics and
  Photonics}\ (\bibinfo  {publisher} {SPIE},\ \bibinfo {year} {1990})\ pp.\
  \bibinfo {pages} {223 -- 237}\BibitemShut {NoStop}%
\bibitem [{\citenamefont {Boukhayma}(2017)}]{boukhayma2017ultra}%
  \BibitemOpen
  \bibfield  {author} {\bibinfo {author} {\bibfnamefont {A.}~\bibnamefont
  {Boukhayma}},\ }\href {https://books.google.com.ar/books?id=ippADwAAQBAJ}
  {\emph {\bibinfo {title} {Ultra Low Noise CMOS Image Sensors}}},\ Springer
  Theses\ (\bibinfo  {publisher} {Springer International Publishing},\ \bibinfo
  {year} {2017})\BibitemShut {NoStop}%
\bibitem [{\citenamefont {Hynecek}\ and\ \citenamefont
  {Nishiwaki}(2003)}]{Hynecek2003}%
  \BibitemOpen
  \bibfield  {author} {\bibinfo {author} {\bibfnamefont {J.}~\bibnamefont
  {Hynecek}}\ and\ \bibinfo {author} {\bibfnamefont {T.}~\bibnamefont
  {Nishiwaki}},\ }\href {\doibase 10.1109/TED.2002.806962} {\bibfield
  {journal} {\bibinfo  {journal} {IEEE Transactions on Electron Devices}\
  }\textbf {\bibinfo {volume} {50}},\ \bibinfo {pages} {239} (\bibinfo {year}
  {2003})}\BibitemShut {NoStop}%
\bibitem [{\citenamefont {Buzhan}\ \emph {et~al.}(2003)\citenamefont {Buzhan},
  \citenamefont {Dolgoshein}, \citenamefont {Filatov}, \citenamefont {Ilyin},
  \citenamefont {Kantzerov}, \citenamefont {Kaplin}, \citenamefont {Karakash},
  \citenamefont {Kayumov}, \citenamefont {Klemin}, \citenamefont {Popova},\
  and\ \citenamefont {Smirnov}}]{BUZHAN2003}%
  \BibitemOpen
  \bibfield  {author} {\bibinfo {author} {\bibfnamefont {P.}~\bibnamefont
  {Buzhan}}, \bibinfo {author} {\bibfnamefont {B.}~\bibnamefont {Dolgoshein}},
  \bibinfo {author} {\bibfnamefont {L.}~\bibnamefont {Filatov}}, \bibinfo
  {author} {\bibfnamefont {A.}~\bibnamefont {Ilyin}}, \bibinfo {author}
  {\bibfnamefont {V.}~\bibnamefont {Kantzerov}}, \bibinfo {author}
  {\bibfnamefont {V.}~\bibnamefont {Kaplin}}, \bibinfo {author} {\bibfnamefont
  {A.}~\bibnamefont {Karakash}}, \bibinfo {author} {\bibfnamefont
  {F.}~\bibnamefont {Kayumov}}, \bibinfo {author} {\bibfnamefont
  {S.}~\bibnamefont {Klemin}}, \bibinfo {author} {\bibfnamefont
  {E.}~\bibnamefont {Popova}}, \ and\ \bibinfo {author} {\bibfnamefont
  {S.}~\bibnamefont {Smirnov}},\ }\href {\doibase
  https://doi.org/10.1016/S0168-9002(03)00749-6} {\bibfield  {journal}
  {\bibinfo  {journal} {Nuclear Instruments and Methods in Physics Research
  Section A: Accelerators, Spectrometers, Detectors and Associated Equipment}\
  }\textbf {\bibinfo {volume} {504}},\ \bibinfo {pages} {48} (\bibinfo {year}
  {2003})},\ \bibinfo {note} {proceedings of the 3rd International Conference
  on New Developments in Photodetection}\BibitemShut {NoStop}%
\bibitem [{\citenamefont {Fossum}\ \emph {et~al.}(2016)\citenamefont {Fossum},
  \citenamefont {Ma}, \citenamefont {Masoodian}, \citenamefont {Anzagira},\
  and\ \citenamefont {Zizza}}]{Fossum_2016}%
  \BibitemOpen
  \bibfield  {author} {\bibinfo {author} {\bibfnamefont {E.~R.}\ \bibnamefont
  {Fossum}}, \bibinfo {author} {\bibfnamefont {J.}~\bibnamefont {Ma}}, \bibinfo
  {author} {\bibfnamefont {S.}~\bibnamefont {Masoodian}}, \bibinfo {author}
  {\bibfnamefont {L.}~\bibnamefont {Anzagira}}, \ and\ \bibinfo {author}
  {\bibfnamefont {R.}~\bibnamefont {Zizza}},\ }\href {\doibase
  10.3390/s16081260} {\bibfield  {journal} {\bibinfo  {journal} {Sensors}\
  }\textbf {\bibinfo {volume} {16}} (\bibinfo {year} {2016}),\
  10.3390/s16081260}\BibitemShut {NoStop}%
\bibitem [{\citenamefont {Wen}\ and\ \citenamefont
  {Salsbury}(1973)}]{Wen_1973}%
  \BibitemOpen
  \bibfield  {author} {\bibinfo {author} {\bibfnamefont {D.~D.}\ \bibnamefont
  {Wen}}\ and\ \bibinfo {author} {\bibfnamefont {P.~J.}\ \bibnamefont
  {Salsbury}},\ }\href@noop {} {\bibfield  {journal} {\bibinfo  {journal}
  {ISSCC Dig: Tech. Papers}\ ,\ \bibinfo {pages} {154}} (\bibinfo {year}
  {1973})}\BibitemShut {NoStop}%
\bibitem [{\citenamefont {Wen}(1974)}]{Wen_1974}%
  \BibitemOpen
  \bibfield  {author} {\bibinfo {author} {\bibfnamefont {D.}~\bibnamefont
  {Wen}},\ }\href {\doibase 10.1109/JSSC.1974.1050535} {\bibfield  {journal}
  {\bibinfo  {journal} {IEEE Journal of Solid-State Circuits}\ }\textbf
  {\bibinfo {volume} {9}},\ \bibinfo {pages} {410} (\bibinfo {year}
  {1974})}\BibitemShut {NoStop}%
\bibitem [{\citenamefont {Wen}\ \emph {et~al.}(1975)\citenamefont {Wen},
  \citenamefont {Early}, \citenamefont {Kim},\ and\ \citenamefont
  {Amelio}}]{FG_1975}%
  \BibitemOpen
  \bibfield  {author} {\bibinfo {author} {\bibfnamefont {D.}~\bibnamefont
  {Wen}}, \bibinfo {author} {\bibfnamefont {J.}~\bibnamefont {Early}}, \bibinfo
  {author} {\bibfnamefont {C.}~\bibnamefont {Kim}}, \ and\ \bibinfo {author}
  {\bibfnamefont {G.}~\bibnamefont {Amelio}},\ }in\ \href {\doibase
  10.1109/ISSCC.1975.1155402} {\emph {\bibinfo {booktitle} {1975 IEEE
  International Solid-State Circuits Conference. Digest of Technical
  Papers}}},\ Vol.\ \bibinfo {volume} {XVIII}\ (\bibinfo {year} {1975})\ pp.\
  \bibinfo {pages} {24--25}\BibitemShut {NoStop}%
\bibitem [{\citenamefont {Janesick}(1993)}]{Janesick_patent}%
  \BibitemOpen
  \bibfield  {author} {\bibinfo {author} {\bibfnamefont {J.~R.}\ \bibnamefont
  {Janesick}},\ }\href@noop {} {\enquote {\bibinfo {title} {Ultra low-noise
  charge coupled device},}\ } (\bibinfo {year} {1993}),\ \Eprint
  {http://arxiv.org/abs/US Patent 5250824A} {US Patent 5250824A} \BibitemShut
  {NoStop}%
\bibitem [{\citenamefont {Chandler}\ \emph {et~al.}(1990)\citenamefont
  {Chandler}, \citenamefont {Bredthauer}, \citenamefont {Janesick},\ and\
  \citenamefont {Westphal}}]{Chandler_1990}%
  \BibitemOpen
  \bibfield  {author} {\bibinfo {author} {\bibfnamefont {C.~E.}\ \bibnamefont
  {Chandler}}, \bibinfo {author} {\bibfnamefont {R.~A.}\ \bibnamefont
  {Bredthauer}}, \bibinfo {author} {\bibfnamefont {J.~R.}\ \bibnamefont
  {Janesick}}, \ and\ \bibinfo {author} {\bibfnamefont {J.~A.}\ \bibnamefont
  {Westphal}},\ }in\ \href {\doibase 10.1117/12.19457} {\emph {\bibinfo
  {booktitle} {Charge-Coupled Devices and Solid State Optical Sensors}}},\
  Vol.\ \bibinfo {volume} {1242},\ \bibinfo {editor} {edited by\ \bibinfo
  {editor} {\bibfnamefont {M.~M.}\ \bibnamefont {Blouke}}},\ \bibinfo
  {organization} {International Society for Optics and Photonics}\ (\bibinfo
  {publisher} {SPIE},\ \bibinfo {year} {1990})\ pp.\ \bibinfo {pages} {238 --
  251}\BibitemShut {NoStop}%
\bibitem [{\citenamefont {Hynecek}(1997)}]{Hynecek_1997}%
  \BibitemOpen
  \bibfield  {author} {\bibinfo {author} {\bibfnamefont {J.}~\bibnamefont
  {Hynecek}},\ }\href {\doibase 10.1109/16.628823} {\bibfield  {journal}
  {\bibinfo  {journal} {IEEE Transactions on Electron Devices}\ }\textbf
  {\bibinfo {volume} {44}},\ \bibinfo {pages} {1679} (\bibinfo {year}
  {1997})}\BibitemShut {NoStop}%
\bibitem [{\citenamefont {Holland}\ \emph {et~al.}(2003)\citenamefont
  {Holland}, \citenamefont {Groom}, \citenamefont {Palaio}, \citenamefont
  {Stover},\ and\ \citenamefont {Wei}}]{Holland:2003}%
  \BibitemOpen
  \bibfield  {author} {\bibinfo {author} {\bibfnamefont {S.~E.}\ \bibnamefont
  {Holland}}, \bibinfo {author} {\bibfnamefont {D.~E.}\ \bibnamefont {Groom}},
  \bibinfo {author} {\bibfnamefont {N.~P.}\ \bibnamefont {Palaio}}, \bibinfo
  {author} {\bibfnamefont {R.~J.}\ \bibnamefont {Stover}}, \ and\ \bibinfo
  {author} {\bibfnamefont {M.}~\bibnamefont {Wei}},\ }\href {\doibase
  10.1109/TED.2002.806476} {\bibfield  {journal} {\bibinfo  {journal} {IEEE
  Transactions on Electron Devices}\ }\textbf {\bibinfo {volume} {50}},\
  \bibinfo {pages} {225} (\bibinfo {year} {2003})}\BibitemShut {NoStop}%
\bibitem [{\citenamefont {Holland}\ \emph {et~al.}(2006)\citenamefont
  {Holland}, \citenamefont {Bebek}, \citenamefont {Dawson}, \citenamefont
  {Emes}, \citenamefont {Fabricius}, \citenamefont {Fairfield}, \citenamefont
  {Groom}, \citenamefont {Karcher}, \citenamefont {Kolbe}, \citenamefont
  {Palaio}, \citenamefont {Roe},\ and\ \citenamefont {Wang}}]{HV_2006}%
  \BibitemOpen
  \bibfield  {author} {\bibinfo {author} {\bibfnamefont {S.~E.}\ \bibnamefont
  {Holland}}, \bibinfo {author} {\bibfnamefont {C.~J.}\ \bibnamefont {Bebek}},
  \bibinfo {author} {\bibfnamefont {K.~S.}\ \bibnamefont {Dawson}}, \bibinfo
  {author} {\bibfnamefont {J.~H.}\ \bibnamefont {Emes}}, \bibinfo {author}
  {\bibfnamefont {M.~H.}\ \bibnamefont {Fabricius}}, \bibinfo {author}
  {\bibfnamefont {J.~A.}\ \bibnamefont {Fairfield}}, \bibinfo {author}
  {\bibfnamefont {D.~E.}\ \bibnamefont {Groom}}, \bibinfo {author}
  {\bibfnamefont {A.}~\bibnamefont {Karcher}}, \bibinfo {author} {\bibfnamefont
  {W.~F.}\ \bibnamefont {Kolbe}}, \bibinfo {author} {\bibfnamefont {N.~P.}\
  \bibnamefont {Palaio}}, \bibinfo {author} {\bibfnamefont {N.~A.}\
  \bibnamefont {Roe}}, \ and\ \bibinfo {author} {\bibfnamefont
  {G.}~\bibnamefont {Wang}},\ }in\ \href {\doibase 10.1117/12.672393} {\emph
  {\bibinfo {booktitle} {High Energy, Optical, and Infrared Detectors for
  Astronomy II}}},\ Vol.\ \bibinfo {volume} {6276},\ \bibinfo {editor} {edited
  by\ \bibinfo {editor} {\bibfnamefont {D.~A.}\ \bibnamefont {Dorn}}\ and\
  \bibinfo {editor} {\bibfnamefont {A.~D.}\ \bibnamefont {Holland}}},\ \bibinfo
  {organization} {International Society for Optics and Photonics}\ (\bibinfo
  {publisher} {SPIE},\ \bibinfo {year} {2006})\ p.\ \bibinfo {pages}
  {62760B}\BibitemShut {NoStop}%
\bibitem [{\citenamefont {Holland}\ \emph {et~al.}(2009)\citenamefont
  {Holland}, \citenamefont {Kolbe},\ and\ \citenamefont {Bebek}}]{HV_2009}%
  \BibitemOpen
  \bibfield  {author} {\bibinfo {author} {\bibfnamefont {S.~E.}\ \bibnamefont
  {Holland}}, \bibinfo {author} {\bibfnamefont {W.~F.}\ \bibnamefont {Kolbe}},
  \ and\ \bibinfo {author} {\bibfnamefont {C.~J.}\ \bibnamefont {Bebek}},\
  }\href {\doibase 10.1109/TED.2009.2030631} {\bibfield  {journal} {\bibinfo
  {journal} {IEEE Transactions on Electron Devices}\ }\textbf {\bibinfo
  {volume} {56}},\ \bibinfo {pages} {2612} (\bibinfo {year}
  {2009})}\BibitemShut {NoStop}%
\bibitem [{\citenamefont {Fernandez~Moroni}\ \emph {et~al.}(2012)\citenamefont
  {Fernandez~Moroni}, \citenamefont {Estrada}, \citenamefont {Cancelo},
  \citenamefont {Holland}, \citenamefont {Paolini},\ and\ \citenamefont
  {Diehl}}]{skipper_2012}%
  \BibitemOpen
  \bibfield  {author} {\bibinfo {author} {\bibfnamefont {G.}~\bibnamefont
  {Fernandez~Moroni}}, \bibinfo {author} {\bibfnamefont {J.}~\bibnamefont
  {Estrada}}, \bibinfo {author} {\bibfnamefont {G.}~\bibnamefont {Cancelo}},
  \bibinfo {author} {\bibfnamefont {S.}~\bibnamefont {Holland}}, \bibinfo
  {author} {\bibfnamefont {E.}~\bibnamefont {Paolini}}, \ and\ \bibinfo
  {author} {\bibfnamefont {H.}~\bibnamefont {Diehl}},\ }\href {\doibase
  10.1007/s10686-012-9298-x} {\bibfield  {journal} {\bibinfo  {journal}
  {Experimental Astronomy}\ }\textbf {\bibinfo {volume} {34}} (\bibinfo {year}
  {2012}),\ 10.1007/s10686-012-9298-x}\BibitemShut {NoStop}%
\bibitem [{\citenamefont {Tiffenberg}\ \emph {et~al.}(2017)\citenamefont
  {Tiffenberg}, \citenamefont {Sofo-Haro}, \citenamefont {Drlica-Wagner},
  \citenamefont {Essig}, \citenamefont {Guardincerri}, \citenamefont {Holland},
  \citenamefont {Volansky},\ and\ \citenamefont {Yu}}]{Tiffenberg:2017aac}%
  \BibitemOpen
  \bibfield  {author} {\bibinfo {author} {\bibfnamefont {J.}~\bibnamefont
  {Tiffenberg}}, \bibinfo {author} {\bibfnamefont {M.}~\bibnamefont
  {Sofo-Haro}}, \bibinfo {author} {\bibfnamefont {A.}~\bibnamefont
  {Drlica-Wagner}}, \bibinfo {author} {\bibfnamefont {R.}~\bibnamefont
  {Essig}}, \bibinfo {author} {\bibfnamefont {Y.}~\bibnamefont {Guardincerri}},
  \bibinfo {author} {\bibfnamefont {S.}~\bibnamefont {Holland}}, \bibinfo
  {author} {\bibfnamefont {T.}~\bibnamefont {Volansky}}, \ and\ \bibinfo
  {author} {\bibfnamefont {T.-T.}\ \bibnamefont {Yu}},\ }\href {\doibase
  10.1103/PhysRevLett.119.131802} {\bibfield  {journal} {\bibinfo  {journal}
  {Phys. Rev. Lett.}\ }\textbf {\bibinfo {volume} {119}},\ \bibinfo {pages}
  {131802} (\bibinfo {year} {2017})},\ \Eprint
  {http://arxiv.org/abs/1706.00028} {1706.00028} \BibitemShut {NoStop}%
\bibitem [{\citenamefont {Cancelo}\ \emph {et~al.}(2021)\citenamefont
  {Cancelo}, \citenamefont {Chavez}, \citenamefont {Chierchie}, \citenamefont
  {Estrada}, \citenamefont {Fernandez-Moroni}, \citenamefont {Paolini},
  \citenamefont {Haro}, \citenamefont {Soto}, \citenamefont {Stefanazzi},
  \citenamefont {Tiffenberg}, \citenamefont {Treptow}, \citenamefont {Wilcer},\
  and\ \citenamefont {Zmuda}}]{cancelo2021low}%
  \BibitemOpen
  \bibfield  {author} {\bibinfo {author} {\bibfnamefont {G.~I.}\ \bibnamefont
  {Cancelo}}, \bibinfo {author} {\bibfnamefont {C.}~\bibnamefont {Chavez}},
  \bibinfo {author} {\bibfnamefont {F.}~\bibnamefont {Chierchie}}, \bibinfo
  {author} {\bibfnamefont {J.}~\bibnamefont {Estrada}}, \bibinfo {author}
  {\bibfnamefont {G.}~\bibnamefont {Fernandez-Moroni}}, \bibinfo {author}
  {\bibfnamefont {E.~E.}\ \bibnamefont {Paolini}}, \bibinfo {author}
  {\bibfnamefont {M.~S.}\ \bibnamefont {Haro}}, \bibinfo {author}
  {\bibfnamefont {A.}~\bibnamefont {Soto}}, \bibinfo {author} {\bibfnamefont
  {L.}~\bibnamefont {Stefanazzi}}, \bibinfo {author} {\bibfnamefont
  {J.}~\bibnamefont {Tiffenberg}}, \bibinfo {author} {\bibfnamefont
  {K.}~\bibnamefont {Treptow}}, \bibinfo {author} {\bibfnamefont
  {N.}~\bibnamefont {Wilcer}}, \ and\ \bibinfo {author} {\bibfnamefont {T.~J.}\
  \bibnamefont {Zmuda}},\ }\href {\doibase 10.1117/1.JATIS.7.1.015001}
  {\bibfield  {journal} {\bibinfo  {journal} {Journal of Astronomical
  Telescopes, Instruments, and Systems}\ }\textbf {\bibinfo {volume} {7}},\
  \bibinfo {pages} {1 } (\bibinfo {year} {2021})}\BibitemShut {NoStop}%
\bibitem [{\citenamefont {Barak}\ \emph {et~al.}(2022)\citenamefont {Barak},
  \citenamefont {Bloch}, \citenamefont {Botti}, \citenamefont {Cababie},
  \citenamefont {Cancelo}, \citenamefont {Chaplinsky}, \citenamefont
  {Chierchie}, \citenamefont {Crisler}, \citenamefont {Drlica-Wagner},
  \citenamefont {Essig}, \citenamefont {Estrada}, \citenamefont {Etzion},
  \citenamefont {Fernandez~Moroni}, \citenamefont {Gift}, \citenamefont
  {Holland}, \citenamefont {Munagavalasa}, \citenamefont {Orly}, \citenamefont
  {Rodrigues}, \citenamefont {Singal}, \citenamefont {Haro}, \citenamefont
  {Stefanazzi}, \citenamefont {Tiffenberg}, \citenamefont {Uemura},
  \citenamefont {Volansky},\ and\ \citenamefont {Yu}}]{cababie_2022}%
  \BibitemOpen
  \bibfield  {author} {\bibinfo {author} {\bibfnamefont {L.}~\bibnamefont
  {Barak}}, \bibinfo {author} {\bibfnamefont {I.~M.}\ \bibnamefont {Bloch}},
  \bibinfo {author} {\bibfnamefont {A.}~\bibnamefont {Botti}}, \bibinfo
  {author} {\bibfnamefont {M.}~\bibnamefont {Cababie}}, \bibinfo {author}
  {\bibfnamefont {G.}~\bibnamefont {Cancelo}}, \bibinfo {author} {\bibfnamefont
  {L.}~\bibnamefont {Chaplinsky}}, \bibinfo {author} {\bibfnamefont
  {F.}~\bibnamefont {Chierchie}}, \bibinfo {author} {\bibfnamefont
  {M.}~\bibnamefont {Crisler}}, \bibinfo {author} {\bibfnamefont
  {A.}~\bibnamefont {Drlica-Wagner}}, \bibinfo {author} {\bibfnamefont
  {R.}~\bibnamefont {Essig}}, \bibinfo {author} {\bibfnamefont
  {J.}~\bibnamefont {Estrada}}, \bibinfo {author} {\bibfnamefont
  {E.}~\bibnamefont {Etzion}}, \bibinfo {author} {\bibfnamefont
  {G.}~\bibnamefont {Fernandez~Moroni}}, \bibinfo {author} {\bibfnamefont
  {D.}~\bibnamefont {Gift}}, \bibinfo {author} {\bibfnamefont {S.~E.}\
  \bibnamefont {Holland}}, \bibinfo {author} {\bibfnamefont {S.}~\bibnamefont
  {Munagavalasa}}, \bibinfo {author} {\bibfnamefont {A.}~\bibnamefont {Orly}},
  \bibinfo {author} {\bibfnamefont {D.}~\bibnamefont {Rodrigues}}, \bibinfo
  {author} {\bibfnamefont {A.}~\bibnamefont {Singal}}, \bibinfo {author}
  {\bibfnamefont {M.~S.}\ \bibnamefont {Haro}}, \bibinfo {author}
  {\bibfnamefont {L.}~\bibnamefont {Stefanazzi}}, \bibinfo {author}
  {\bibfnamefont {J.}~\bibnamefont {Tiffenberg}}, \bibinfo {author}
  {\bibfnamefont {S.}~\bibnamefont {Uemura}}, \bibinfo {author} {\bibfnamefont
  {T.}~\bibnamefont {Volansky}}, \ and\ \bibinfo {author} {\bibfnamefont
  {T.-T.}\ \bibnamefont {Yu}} (\bibinfo {collaboration} {SENSEI
  Collaboration}),\ }\href {\doibase 10.1103/PhysRevApplied.17.014022}
  {\bibfield  {journal} {\bibinfo  {journal} {Phys. Rev. Appl.}\ }\textbf
  {\bibinfo {volume} {17}},\ \bibinfo {pages} {014022} (\bibinfo {year}
  {2022})}\BibitemShut {NoStop}%
\bibitem [{\citenamefont {Barak}\ \emph {et~al.}(2020)\citenamefont {Barak},
  \citenamefont {Bloch}, \citenamefont {Cababie}, \citenamefont {Cancelo},
  \citenamefont {Chaplinsky}, \citenamefont {Chierchie}, \citenamefont
  {Crisler}, \citenamefont {Drlica-Wagner}, \citenamefont {Essig},
  \citenamefont {Estrada}, \citenamefont {Etzion}, \citenamefont {Moroni},
  \citenamefont {Gift}, \citenamefont {Munagavalasa}, \citenamefont {Orly},
  \citenamefont {Rodrigues}, \citenamefont {Singal}, \citenamefont {Haro},
  \citenamefont {Stefanazzi}, \citenamefont {Tiffenberg}, \citenamefont
  {Uemura}, \citenamefont {Volansky},\ and\ \citenamefont {Yu}}]{Barak2020}%
  \BibitemOpen
  \bibfield  {author} {\bibinfo {author} {\bibfnamefont {L.}~\bibnamefont
  {Barak}}, \bibinfo {author} {\bibfnamefont {I.~M.}\ \bibnamefont {Bloch}},
  \bibinfo {author} {\bibfnamefont {M.}~\bibnamefont {Cababie}}, \bibinfo
  {author} {\bibfnamefont {G.}~\bibnamefont {Cancelo}}, \bibinfo {author}
  {\bibfnamefont {L.}~\bibnamefont {Chaplinsky}}, \bibinfo {author}
  {\bibfnamefont {F.}~\bibnamefont {Chierchie}}, \bibinfo {author}
  {\bibfnamefont {M.}~\bibnamefont {Crisler}}, \bibinfo {author} {\bibfnamefont
  {A.}~\bibnamefont {Drlica-Wagner}}, \bibinfo {author} {\bibfnamefont
  {R.}~\bibnamefont {Essig}}, \bibinfo {author} {\bibfnamefont
  {J.}~\bibnamefont {Estrada}}, \bibinfo {author} {\bibfnamefont
  {E.}~\bibnamefont {Etzion}}, \bibinfo {author} {\bibfnamefont {G.~F.}\
  \bibnamefont {Moroni}}, \bibinfo {author} {\bibfnamefont {D.}~\bibnamefont
  {Gift}}, \bibinfo {author} {\bibfnamefont {S.}~\bibnamefont {Munagavalasa}},
  \bibinfo {author} {\bibfnamefont {A.}~\bibnamefont {Orly}}, \bibinfo {author}
  {\bibfnamefont {D.}~\bibnamefont {Rodrigues}}, \bibinfo {author}
  {\bibfnamefont {A.}~\bibnamefont {Singal}}, \bibinfo {author} {\bibfnamefont
  {M.~S.}\ \bibnamefont {Haro}}, \bibinfo {author} {\bibfnamefont
  {L.}~\bibnamefont {Stefanazzi}}, \bibinfo {author} {\bibfnamefont
  {J.}~\bibnamefont {Tiffenberg}}, \bibinfo {author} {\bibfnamefont
  {S.}~\bibnamefont {Uemura}}, \bibinfo {author} {\bibfnamefont
  {T.}~\bibnamefont {Volansky}}, \ and\ \bibinfo {author} {\bibfnamefont
  {T.-T.}\ \bibnamefont {Yu}} (\bibinfo {collaboration} {SENSEI
  Collaboration}),\ }\href {\doibase 10.1103/PhysRevLett.125.171802} {\bibfield
   {journal} {\bibinfo  {journal} {Phys. Rev. Lett.}\ }\textbf {\bibinfo
  {volume} {125}},\ \bibinfo {pages} {171802} (\bibinfo {year}
  {2020})}\BibitemShut {NoStop}%
\bibitem [{\citenamefont {Drlica-Wagner}\ \emph {et~al.}(2020)\citenamefont
  {Drlica-Wagner}, \citenamefont {Villalpando}, \citenamefont {O'Neil},
  \citenamefont {Estrada}, \citenamefont {Holland}, \citenamefont {Kurinsky},
  \citenamefont {Li}, \citenamefont {Moroni}, \citenamefont {Tiffenberg},\ and\
  \citenamefont {Uemura}}]{Drlica_2020}%
  \BibitemOpen
  \bibfield  {author} {\bibinfo {author} {\bibfnamefont {A.}~\bibnamefont
  {Drlica-Wagner}}, \bibinfo {author} {\bibfnamefont {E.~M.}\ \bibnamefont
  {Villalpando}}, \bibinfo {author} {\bibfnamefont {J.}~\bibnamefont {O'Neil}},
  \bibinfo {author} {\bibfnamefont {J.}~\bibnamefont {Estrada}}, \bibinfo
  {author} {\bibfnamefont {S.}~\bibnamefont {Holland}}, \bibinfo {author}
  {\bibfnamefont {N.}~\bibnamefont {Kurinsky}}, \bibinfo {author}
  {\bibfnamefont {T.}~\bibnamefont {Li}}, \bibinfo {author} {\bibfnamefont
  {G.~F.}\ \bibnamefont {Moroni}}, \bibinfo {author} {\bibfnamefont
  {J.}~\bibnamefont {Tiffenberg}}, \ and\ \bibinfo {author} {\bibfnamefont
  {S.}~\bibnamefont {Uemura}},\ }in\ \href {\doibase 10.1117/12.2562403} {\emph
  {\bibinfo {booktitle} {X-Ray, Optical, and Infrared Detectors for Astronomy
  IX}}},\ Vol.\ \bibinfo {volume} {11454},\ \bibinfo {editor} {edited by\
  \bibinfo {editor} {\bibfnamefont {A.~D.}\ \bibnamefont {Holland}}\ and\
  \bibinfo {editor} {\bibfnamefont {J.}~\bibnamefont {Beletic}}},\ \bibinfo
  {organization} {International Society for Optics and Photonics}\ (\bibinfo
  {publisher} {SPIE},\ \bibinfo {year} {2020})\ pp.\ \bibinfo {pages} {210 --
  223}\BibitemShut {NoStop}%
\bibitem [{\citenamefont {Estrada}(2021)}]{OSCURA2020}%
  \BibitemOpen
  \bibfield  {author} {\bibinfo {author} {\bibfnamefont {J.}~\bibnamefont
  {Estrada}},\ }\href@noop {} {\emph {\bibinfo {title} {Observatory of
  {Skipper} {CCD}s Unveiling Recoiling Atoms}}} (\bibinfo {year} {2020
  (accessed September 23, 2021)}),\ \bibinfo {note}
  {\url{https://astro.fnal.gov/science/dark-matter/oscura/}}\BibitemShut
  {NoStop}%
\bibitem [{\citenamefont {D'Olivo}\ \emph {et~al.}(2020)\citenamefont
  {D'Olivo}, \citenamefont {Bonifazi}, \citenamefont {Rodrigues},\ and\
  \citenamefont {Moroni}}]{violeta2020}%
  \BibitemOpen
  \bibfield  {author} {\bibinfo {author} {\bibfnamefont {J.~C.}\ \bibnamefont
  {D'Olivo}}, \bibinfo {author} {\bibfnamefont {C.}~\bibnamefont {Bonifazi}},
  \bibinfo {author} {\bibfnamefont {D.}~\bibnamefont {Rodrigues}}, \ and\
  \bibinfo {author} {\bibfnamefont {G.~F.}\ \bibnamefont {Moroni}},\ }in\
  \href@noop {} {\emph {\bibinfo {booktitle} {XXIX International Conference in
  Neutrino Physics, poster 521}}}\ (\bibinfo {year} {2020})\BibitemShut
  {NoStop}%
\bibitem [{\citenamefont {Rodrigues}\ \emph {et~al.}(2020)\citenamefont
  {Rodrigues}, \citenamefont {Andersson}, \citenamefont {Cababie},
  \citenamefont {Donadon}, \citenamefont {Cancelo}, \citenamefont {Estrada},
  \citenamefont {Fernandez-Moroni}, \citenamefont {Piegaia}, \citenamefont
  {Senger}, \citenamefont {Haro} \emph {et~al.}}]{Rodrigues_2020}%
  \BibitemOpen
  \bibfield  {author} {\bibinfo {author} {\bibfnamefont {D.}~\bibnamefont
  {Rodrigues}}, \bibinfo {author} {\bibfnamefont {K.}~\bibnamefont
  {Andersson}}, \bibinfo {author} {\bibfnamefont {M.}~\bibnamefont {Cababie}},
  \bibinfo {author} {\bibfnamefont {A.}~\bibnamefont {Donadon}}, \bibinfo
  {author} {\bibfnamefont {G.}~\bibnamefont {Cancelo}}, \bibinfo {author}
  {\bibfnamefont {J.}~\bibnamefont {Estrada}}, \bibinfo {author} {\bibfnamefont
  {G.}~\bibnamefont {Fernandez-Moroni}}, \bibinfo {author} {\bibfnamefont
  {R.}~\bibnamefont {Piegaia}}, \bibinfo {author} {\bibfnamefont
  {M.}~\bibnamefont {Senger}}, \bibinfo {author} {\bibfnamefont {M.~S.}\
  \bibnamefont {Haro}},  \emph {et~al.},\ }\href@noop {} {\bibfield  {journal}
  {\bibinfo  {journal} {arXiv preprint arXiv:2004.11499}\ } (\bibinfo {year}
  {2020})}\BibitemShut {NoStop}%
\bibitem [{\citenamefont {Botti}\ \emph {et~al.}(2022)\citenamefont {Botti},
  \citenamefont {Uemura}, \citenamefont {Moroni}, \citenamefont {Barak},
  \citenamefont {Cababie}, \citenamefont {Essig}, \citenamefont {Etzion},
  \citenamefont {Rodrigues}, \citenamefont {Saffold}, \citenamefont
  {Sofo~Haro}, \citenamefont {Tiffenberg},\ and\ \citenamefont
  {Volansky}}]{botti_2022}%
  \BibitemOpen
  \bibfield  {author} {\bibinfo {author} {\bibfnamefont {A.~M.}\ \bibnamefont
  {Botti}}, \bibinfo {author} {\bibfnamefont {S.}~\bibnamefont {Uemura}},
  \bibinfo {author} {\bibfnamefont {G.~F.}\ \bibnamefont {Moroni}}, \bibinfo
  {author} {\bibfnamefont {L.}~\bibnamefont {Barak}}, \bibinfo {author}
  {\bibfnamefont {M.}~\bibnamefont {Cababie}}, \bibinfo {author} {\bibfnamefont
  {R.}~\bibnamefont {Essig}}, \bibinfo {author} {\bibfnamefont
  {E.}~\bibnamefont {Etzion}}, \bibinfo {author} {\bibfnamefont
  {D.}~\bibnamefont {Rodrigues}}, \bibinfo {author} {\bibfnamefont
  {N.}~\bibnamefont {Saffold}}, \bibinfo {author} {\bibfnamefont
  {M.}~\bibnamefont {Sofo~Haro}}, \bibinfo {author} {\bibfnamefont
  {J.}~\bibnamefont {Tiffenberg}}, \ and\ \bibinfo {author} {\bibfnamefont
  {T.}~\bibnamefont {Volansky}},\ }\href {\doibase 10.1103/PhysRevD.106.072005}
  {\bibfield  {journal} {\bibinfo  {journal} {Phys. Rev. D}\ }\textbf {\bibinfo
  {volume} {106}},\ \bibinfo {pages} {072005} (\bibinfo {year}
  {2022})}\BibitemShut {NoStop}%
\bibitem [{\citenamefont {Estrada}\ \emph {et~al.}(2021)\citenamefont
  {Estrada}, \citenamefont {Harnik}, \citenamefont {Rodrigues},\ and\
  \citenamefont {Senger}}]{estrada2021ghost}%
  \BibitemOpen
  \bibfield  {author} {\bibinfo {author} {\bibfnamefont {J.}~\bibnamefont
  {Estrada}}, \bibinfo {author} {\bibfnamefont {R.}~\bibnamefont {Harnik}},
  \bibinfo {author} {\bibfnamefont {D.}~\bibnamefont {Rodrigues}}, \ and\
  \bibinfo {author} {\bibfnamefont {M.}~\bibnamefont {Senger}},\ }\href@noop {}
  {\enquote {\bibinfo {title} {Ghost imaging of dark particles},}\ } (\bibinfo
  {year} {2021}),\ \Eprint {http://arxiv.org/abs/2012.04707} {arXiv:2012.04707
  [hep-ph]} \BibitemShut {NoStop}%
\bibitem [{\citenamefont {Rauscher}\ \emph {et~al.}(2019)\citenamefont
  {Rauscher}, \citenamefont {Holland}, \citenamefont {Miko},\ and\
  \citenamefont {Waczynski}}]{RauscherNASA2019}%
  \BibitemOpen
  \bibfield  {author} {\bibinfo {author} {\bibfnamefont {B.~J.}\ \bibnamefont
  {Rauscher}}, \bibinfo {author} {\bibfnamefont {S.~E.}\ \bibnamefont
  {Holland}}, \bibinfo {author} {\bibfnamefont {L.~R.}\ \bibnamefont {Miko}}, \
  and\ \bibinfo {author} {\bibfnamefont {A.}~\bibnamefont {Waczynski}},\ }in\
  \href {\doibase 10.1117/12.2536509} {\emph {\bibinfo {booktitle}
  {UV/Optical/IR Space Telescopes and Instruments: Innovative Technologies and
  Concepts IX}}},\ Vol.\ \bibinfo {volume} {11115},\ \bibinfo {editor} {edited
  by\ \bibinfo {editor} {\bibfnamefont {A.~A.}\ \bibnamefont {Barto}}, \bibinfo
  {editor} {\bibfnamefont {J.~B.}\ \bibnamefont {Breckinridge}}, \ and\
  \bibinfo {editor} {\bibfnamefont {H.~P.}\ \bibnamefont {Stahl}}},\ \bibinfo
  {organization} {International Society for Optics and Photonics}\ (\bibinfo
  {publisher} {SPIE},\ \bibinfo {year} {2019})\ pp.\ \bibinfo {pages} {382 --
  386}\BibitemShut {NoStop}%
\bibitem [{\citenamefont {Samantaray}\ \emph {et~al.}(2017)\citenamefont
  {Samantaray}, \citenamefont {Ruo-Berchera}, \citenamefont {Meda},\ and\
  \citenamefont {Genovese}}]{Samantaray2017}%
  \BibitemOpen
  \bibfield  {author} {\bibinfo {author} {\bibfnamefont {N.}~\bibnamefont
  {Samantaray}}, \bibinfo {author} {\bibfnamefont {I.}~\bibnamefont
  {Ruo-Berchera}}, \bibinfo {author} {\bibfnamefont {A.}~\bibnamefont {Meda}},
  \ and\ \bibinfo {author} {\bibfnamefont {M.}~\bibnamefont {Genovese}},\
  }\href {\doibase 10.1038/lsa.2017.5} {\bibfield  {journal} {\bibinfo
  {journal} {Light: Science {\&} Applications}\ }\textbf {\bibinfo {volume}
  {6}},\ \bibinfo {pages} {e17005} (\bibinfo {year} {2017})}\BibitemShut
  {NoStop}%
\bibitem [{\citenamefont {Holland}()}]{holland_2023}%
  \BibitemOpen
  \bibfield  {author} {\bibinfo {author} {\bibfnamefont {S.~E.}\ \bibnamefont
  {Holland}},\ }\href {\doibase https://doi.org/10.1002/asna.20230072}
  {\bibfield  {journal} {\bibinfo  {journal} {Astronomische Nachrichten}\
  }\textbf {\bibinfo {volume} {n/a}},\ \bibinfo {pages} {e20230072}},\ \Eprint
  {http://arxiv.org/abs/https://onlinelibrary.wiley.com/doi/pdf/10.1002/asna.20230072}
  {https://onlinelibrary.wiley.com/doi/pdf/10.1002/asna.20230072} \BibitemShut
  {NoStop}%
\bibitem [{\citenamefont {Holland}\ \emph {et~al.}(2007)\citenamefont
  {Holland}, \citenamefont {Dawson}, \citenamefont {Palaio}, \citenamefont
  {Saha}, \citenamefont {Roe},\ and\ \citenamefont {Wang}}]{HOLLAND2007653}%
  \BibitemOpen
  \bibfield  {author} {\bibinfo {author} {\bibfnamefont {S.}~\bibnamefont
  {Holland}}, \bibinfo {author} {\bibfnamefont {K.}~\bibnamefont {Dawson}},
  \bibinfo {author} {\bibfnamefont {N.}~\bibnamefont {Palaio}}, \bibinfo
  {author} {\bibfnamefont {J.}~\bibnamefont {Saha}}, \bibinfo {author}
  {\bibfnamefont {N.}~\bibnamefont {Roe}}, \ and\ \bibinfo {author}
  {\bibfnamefont {G.}~\bibnamefont {Wang}},\ }\href {\doibase
  https://doi.org/10.1016/j.nima.2007.05.265} {\bibfield  {journal} {\bibinfo
  {journal} {Nuclear Instruments and Methods in Physics Research Section A:
  Accelerators, Spectrometers, Detectors and Associated Equipment}\ }\textbf
  {\bibinfo {volume} {579}},\ \bibinfo {pages} {653} (\bibinfo {year}
  {2007})},\ \bibinfo {note} {proceedings of the 6th "Hiroshima" Symposium on
  the Development and Application of Semiconductor Detectors}\BibitemShut
  {NoStop}%
\bibitem [{\citenamefont {Janesick}(2001)}]{janesick2001scientific}%
  \BibitemOpen
  \bibfield  {author} {\bibinfo {author} {\bibfnamefont {J.~R.}\ \bibnamefont
  {Janesick}},\ }\href@noop {} {\emph {\bibinfo {title} {Scientific
  charge-coupled devices}}},\ Vol.~\bibinfo {volume} {83}\ (\bibinfo
  {publisher} {SPIE press},\ \bibinfo {year} {2001})\BibitemShut {NoStop}%
\bibitem [{\citenamefont {{Maruffo Villalpando}}\ \emph
  {et~al.}(2022)\citenamefont {{Maruffo Villalpando}}, \citenamefont
  {{Drlica-Wagner}}, \citenamefont {{Bonati}} \emph {et~al.}}]{Maruffo_2022}%
  \BibitemOpen
  \bibfield  {author} {\bibinfo {author} {\bibfnamefont {E.}~\bibnamefont
  {{Maruffo Villalpando}}}, \bibinfo {author} {\bibfnamefont {A.}~\bibnamefont
  {{Drlica-Wagner}}}, \bibinfo {author} {\bibfnamefont {M.}~\bibnamefont
  {{Bonati}}},  \emph {et~al.},\ }\bibfield  {booktitle} {\emph {\bibinfo
  {booktitle} {{Proceedings, SPIE Astronomical Telescopes + Instrumentation
  2020: X-Ray, Optical, and Infrared Detectors for Astronomy X}}},\ }\href
  {\doibase 10.1117/12.2629475} {\bibfield  {journal} {\bibinfo  {journal}
  {Proc. SPIE Int. Soc. Opt. Eng.}\ }\textbf {\bibinfo {volume} {1219O}},\
  \bibinfo {pages} {12190U} (\bibinfo {year} {2022})},\ \Eprint
  {http://arxiv.org/abs/2210.03665} {arXiv:2210.03665 [astro-ph.IM]}
  \BibitemShut {NoStop}%
\bibitem [{\citenamefont {{National Academies of Sciences, Engineering, and
  Medicine}}(2021)}]{Astro2020}%
  \BibitemOpen
  \bibfield  {author} {\bibinfo {author} {\bibnamefont {{National Academies of
  Sciences, Engineering, and Medicine}}},\ }\href {\doibase 10.17226/26141}
  {\emph {\bibinfo {title} {{Pathways to Discovery in Astronomy and
  Astrophysics for the 2020s}}}}\ (\bibinfo {year} {2021})\BibitemShut
  {NoStop}%
\bibitem [{\citenamefont {Crill}\ and\ \citenamefont
  {Siegler}(2018)}]{crill2018exoplanet}%
  \BibitemOpen
  \bibfield  {author} {\bibinfo {author} {\bibfnamefont {B.}~\bibnamefont
  {Crill}}\ and\ \bibinfo {author} {\bibfnamefont {N.}~\bibnamefont
  {Siegler}},\ }\href@noop {} {\bibfield  {journal} {\bibinfo  {journal} {Jet
  Propulsion Laboratory Publications D-102506}\ } (\bibinfo {year}
  {2018})}\BibitemShut {NoStop}%
\bibitem [{\citenamefont {{Dawson}}\ \emph {et~al.}(2008)\citenamefont
  {{Dawson}}, \citenamefont {{Bebek}}, \citenamefont {{Emes}}, \citenamefont
  {{Holland}}, \citenamefont {{Jelinsky}}, \citenamefont {{Karcher}},
  \citenamefont {{Kolbe}}, \citenamefont {{Palaio}}, \citenamefont {{Roe}},
  \citenamefont {{Saha}}, \citenamefont {{Takasaki}},\ and\ \citenamefont
  {{Wang}}}]{Dawson:2008}%
  \BibitemOpen
  \bibfield  {author} {\bibinfo {author} {\bibfnamefont {K.}~\bibnamefont
  {{Dawson}}}, \bibinfo {author} {\bibfnamefont {C.}~\bibnamefont {{Bebek}}},
  \bibinfo {author} {\bibfnamefont {J.}~\bibnamefont {{Emes}}}, \bibinfo
  {author} {\bibfnamefont {S.}~\bibnamefont {{Holland}}}, \bibinfo {author}
  {\bibfnamefont {S.}~\bibnamefont {{Jelinsky}}}, \bibinfo {author}
  {\bibfnamefont {A.}~\bibnamefont {{Karcher}}}, \bibinfo {author}
  {\bibfnamefont {W.}~\bibnamefont {{Kolbe}}}, \bibinfo {author} {\bibfnamefont
  {N.}~\bibnamefont {{Palaio}}}, \bibinfo {author} {\bibfnamefont
  {N.}~\bibnamefont {{Roe}}}, \bibinfo {author} {\bibfnamefont
  {J.}~\bibnamefont {{Saha}}}, \bibinfo {author} {\bibfnamefont
  {K.}~\bibnamefont {{Takasaki}}}, \ and\ \bibinfo {author} {\bibfnamefont
  {G.}~\bibnamefont {{Wang}}},\ }\href {\doibase 10.1109/TNS.2008.919262}
  {\bibfield  {journal} {\bibinfo  {journal} {IEEE Transactions on Nuclear
  Science}\ }\textbf {\bibinfo {volume} {55}},\ \bibinfo {pages} {1725}
  (\bibinfo {year} {2008})},\ \Eprint {http://arxiv.org/abs/0711.2105}
  {arXiv:0711.2105 [astro-ph]} \BibitemShut {NoStop}%
\bibitem [{\citenamefont {Rauscher}\ \emph {et~al.}()\citenamefont {Rauscher},
  \citenamefont {Holland}, \citenamefont {Miko},\ and\ \citenamefont
  {Waczynski}}]{Rauscher_internal_2019}%
  \BibitemOpen
  \bibfield  {author} {\bibinfo {author} {\bibfnamefont {B.~J.}\ \bibnamefont
  {Rauscher}}, \bibinfo {author} {\bibfnamefont {S.~E.}\ \bibnamefont
  {Holland}}, \bibinfo {author} {\bibfnamefont {L.~R.}\ \bibnamefont {Miko}}, \
  and\ \bibinfo {author} {\bibfnamefont {A.}~\bibnamefont {Waczynski}},\
  }\href@noop {} {\ }\BibitemShut {NoStop}%
\bibitem [{\citenamefont {{Schlegel}}\ \emph {et~al.}(2022)\citenamefont
  {{Schlegel}}, \citenamefont {{Ferraro}}, \citenamefont {{Aldering}},
  \citenamefont {{Baltay}}, \citenamefont {{BenZvi}}, \citenamefont
  {{Besuner}}, \citenamefont {{Blanc}}, \citenamefont {{Bolton}}, \citenamefont
  {{Bonaca}}, \citenamefont {{Brooks}}, \citenamefont {{Buckley-Geer}},
  \citenamefont {{Cai}}, \citenamefont {{DeRose}}, \citenamefont {{Dey}},
  \citenamefont {{Doel}}, \citenamefont {{Drlica-Wagner}}, \citenamefont
  {{Fan}}, \citenamefont {{Gutierrez}}, \citenamefont {{Green}}, \citenamefont
  {{Guy}}, \citenamefont {{Huterer}}, \citenamefont {{Infante}}, \citenamefont
  {{Jelinsky}}, \citenamefont {{Karagiannis}}, \citenamefont {{Kent}},
  \citenamefont {{Kim}}, \citenamefont {{Kneib}}, \citenamefont {{Kollmeier}},
  \citenamefont {{Kremin}}, \citenamefont {{Lahav}}, \citenamefont
  {{Landriau}}, \citenamefont {{Lang}}, \citenamefont {{Leauthaud}},
  \citenamefont {{Levi}}, \citenamefont {{Linder}}, \citenamefont
  {{Magneville}}, \citenamefont {{Martini}}, \citenamefont {{McDonald}},
  \citenamefont {{Miller}}, \citenamefont {{Myers}}, \citenamefont {{Newman}},
  \citenamefont {{Nugent}}, \citenamefont {{Palanque-Delabrouille}},
  \citenamefont {{Padmanabhan}}, \citenamefont {{Palmese}}, \citenamefont
  {{Poppett}}, \citenamefont {{Prochaska}}, \citenamefont {{Raichoor}},
  \citenamefont {{Ramirez}}, \citenamefont {{Sailer}}, \citenamefont
  {{Schaan}}, \citenamefont {{Schubnell}}, \citenamefont {{Seljak}},
  \citenamefont {{Seo}}, \citenamefont {{Silber}}, \citenamefont {{Simon}},
  \citenamefont {{Slepian}}, \citenamefont {{Soares-Santos}}, \citenamefont
  {{Tarle}}, \citenamefont {{Valluri}}, \citenamefont {{Weaverdyck}},
  \citenamefont {{Wechsler}}, \citenamefont {{White}}, \citenamefont
  {{Yeche}},\ and\ \citenamefont {{Zhou}}}]{Schlegel:2022}%
  \BibitemOpen
  \bibfield  {author} {\bibinfo {author} {\bibfnamefont {D.~J.}\ \bibnamefont
  {{Schlegel}}}, \bibinfo {author} {\bibfnamefont {S.}~\bibnamefont
  {{Ferraro}}}, \bibinfo {author} {\bibfnamefont {G.}~\bibnamefont
  {{Aldering}}}, \bibinfo {author} {\bibfnamefont {C.}~\bibnamefont
  {{Baltay}}}, \bibinfo {author} {\bibfnamefont {S.}~\bibnamefont {{BenZvi}}},
  \bibinfo {author} {\bibfnamefont {R.}~\bibnamefont {{Besuner}}}, \bibinfo
  {author} {\bibfnamefont {G.~A.}\ \bibnamefont {{Blanc}}}, \bibinfo {author}
  {\bibfnamefont {A.~S.}\ \bibnamefont {{Bolton}}}, \bibinfo {author}
  {\bibfnamefont {A.}~\bibnamefont {{Bonaca}}}, \bibinfo {author}
  {\bibfnamefont {D.}~\bibnamefont {{Brooks}}}, \bibinfo {author}
  {\bibfnamefont {E.}~\bibnamefont {{Buckley-Geer}}}, \bibinfo {author}
  {\bibfnamefont {Z.}~\bibnamefont {{Cai}}}, \bibinfo {author} {\bibfnamefont
  {J.}~\bibnamefont {{DeRose}}}, \bibinfo {author} {\bibfnamefont
  {A.}~\bibnamefont {{Dey}}}, \bibinfo {author} {\bibfnamefont
  {P.}~\bibnamefont {{Doel}}}, \bibinfo {author} {\bibfnamefont
  {A.}~\bibnamefont {{Drlica-Wagner}}}, \bibinfo {author} {\bibfnamefont
  {X.}~\bibnamefont {{Fan}}}, \bibinfo {author} {\bibfnamefont
  {G.}~\bibnamefont {{Gutierrez}}}, \bibinfo {author} {\bibfnamefont
  {D.}~\bibnamefont {{Green}}}, \bibinfo {author} {\bibfnamefont
  {J.}~\bibnamefont {{Guy}}}, \bibinfo {author} {\bibfnamefont
  {D.}~\bibnamefont {{Huterer}}}, \bibinfo {author} {\bibfnamefont
  {L.}~\bibnamefont {{Infante}}}, \bibinfo {author} {\bibfnamefont
  {P.}~\bibnamefont {{Jelinsky}}}, \bibinfo {author} {\bibfnamefont
  {D.}~\bibnamefont {{Karagiannis}}}, \bibinfo {author} {\bibfnamefont {S.~M.}\
  \bibnamefont {{Kent}}}, \bibinfo {author} {\bibfnamefont {A.~G.}\
  \bibnamefont {{Kim}}}, \bibinfo {author} {\bibfnamefont {J.-P.}\ \bibnamefont
  {{Kneib}}}, \bibinfo {author} {\bibfnamefont {J.~A.}\ \bibnamefont
  {{Kollmeier}}}, \bibinfo {author} {\bibfnamefont {A.}~\bibnamefont
  {{Kremin}}}, \bibinfo {author} {\bibfnamefont {O.}~\bibnamefont {{Lahav}}},
  \bibinfo {author} {\bibfnamefont {M.}~\bibnamefont {{Landriau}}}, \bibinfo
  {author} {\bibfnamefont {D.}~\bibnamefont {{Lang}}}, \bibinfo {author}
  {\bibfnamefont {A.}~\bibnamefont {{Leauthaud}}}, \bibinfo {author}
  {\bibfnamefont {M.~E.}\ \bibnamefont {{Levi}}}, \bibinfo {author}
  {\bibfnamefont {E.~V.}\ \bibnamefont {{Linder}}}, \bibinfo {author}
  {\bibfnamefont {C.}~\bibnamefont {{Magneville}}}, \bibinfo {author}
  {\bibfnamefont {P.}~\bibnamefont {{Martini}}}, \bibinfo {author}
  {\bibfnamefont {P.}~\bibnamefont {{McDonald}}}, \bibinfo {author}
  {\bibfnamefont {C.~J.}\ \bibnamefont {{Miller}}}, \bibinfo {author}
  {\bibfnamefont {A.~D.}\ \bibnamefont {{Myers}}}, \bibinfo {author}
  {\bibfnamefont {J.~A.}\ \bibnamefont {{Newman}}}, \bibinfo {author}
  {\bibfnamefont {P.~E.}\ \bibnamefont {{Nugent}}}, \bibinfo {author}
  {\bibfnamefont {N.}~\bibnamefont {{Palanque-Delabrouille}}}, \bibinfo
  {author} {\bibfnamefont {N.}~\bibnamefont {{Padmanabhan}}}, \bibinfo {author}
  {\bibfnamefont {A.}~\bibnamefont {{Palmese}}}, \bibinfo {author}
  {\bibfnamefont {C.}~\bibnamefont {{Poppett}}}, \bibinfo {author}
  {\bibfnamefont {J.~X.}\ \bibnamefont {{Prochaska}}}, \bibinfo {author}
  {\bibfnamefont {A.}~\bibnamefont {{Raichoor}}}, \bibinfo {author}
  {\bibfnamefont {S.}~\bibnamefont {{Ramirez}}}, \bibinfo {author}
  {\bibfnamefont {N.}~\bibnamefont {{Sailer}}}, \bibinfo {author}
  {\bibfnamefont {E.}~\bibnamefont {{Schaan}}}, \bibinfo {author}
  {\bibfnamefont {M.}~\bibnamefont {{Schubnell}}}, \bibinfo {author}
  {\bibfnamefont {U.}~\bibnamefont {{Seljak}}}, \bibinfo {author}
  {\bibfnamefont {H.-J.}\ \bibnamefont {{Seo}}}, \bibinfo {author}
  {\bibfnamefont {J.}~\bibnamefont {{Silber}}}, \bibinfo {author}
  {\bibfnamefont {J.~D.}\ \bibnamefont {{Simon}}}, \bibinfo {author}
  {\bibfnamefont {Z.}~\bibnamefont {{Slepian}}}, \bibinfo {author}
  {\bibfnamefont {M.}~\bibnamefont {{Soares-Santos}}}, \bibinfo {author}
  {\bibfnamefont {G.}~\bibnamefont {{Tarle}}}, \bibinfo {author} {\bibfnamefont
  {M.}~\bibnamefont {{Valluri}}}, \bibinfo {author} {\bibfnamefont {N.~J.}\
  \bibnamefont {{Weaverdyck}}}, \bibinfo {author} {\bibfnamefont {R.~H.}\
  \bibnamefont {{Wechsler}}}, \bibinfo {author} {\bibfnamefont
  {M.}~\bibnamefont {{White}}}, \bibinfo {author} {\bibfnamefont
  {C.}~\bibnamefont {{Yeche}}}, \ and\ \bibinfo {author} {\bibfnamefont
  {R.}~\bibnamefont {{Zhou}}},\ }\href {\doibase 10.48550/arXiv.2209.03585}
  {\bibfield  {journal} {\bibinfo  {journal} {arXiv e-prints}\ ,\ \bibinfo
  {eid} {arXiv:2209.03585}} (\bibinfo {year} {2022})},\ \Eprint
  {http://arxiv.org/abs/2209.03585} {arXiv:2209.03585 [astro-ph.CO]}
  \BibitemShut {NoStop}%
\bibitem [{\citenamefont {{Asaadi}}\ \emph {et~al.}(2022)\citenamefont
  {{Asaadi}}, \citenamefont {{Baxter}}, \citenamefont {{Berggren}},
  \citenamefont {{Braga}}, \citenamefont {{Charlebois}}, \citenamefont
  {{Chang}}, \citenamefont {{Dragone}}, \citenamefont {{Drlica-Wagner}},
  \citenamefont {{Escobar}}, \citenamefont {{Estrada}}, \citenamefont
  {{Fahim}}, \citenamefont {{Febbraro}}, \citenamefont {{Fernandez Moroni}},
  \citenamefont {{Holland}}, \citenamefont {{Hossbach}}, \citenamefont
  {{Koppell}}, \citenamefont {{Leitz}}, \citenamefont {{Magnoni}},
  \citenamefont {{Mazin}}, \citenamefont {{Pratte}}, \citenamefont
  {{Rauscher}}, \citenamefont {{Rodrigues}}, \citenamefont {{Shen}},
  \citenamefont {{Sofo-Haro}}, \citenamefont {{Tiffenberg}}, \citenamefont
  {{Turner}}, \citenamefont {{Rota}}, \citenamefont {{Kenney}}, \citenamefont
  {{Vachon}},\ and\ \citenamefont {{Wang}}}]{Asaadi:2022}%
  \BibitemOpen
  \bibfield  {author} {\bibinfo {author} {\bibfnamefont {J.}~\bibnamefont
  {{Asaadi}}}, \bibinfo {author} {\bibfnamefont {D.}~\bibnamefont {{Baxter}}},
  \bibinfo {author} {\bibfnamefont {K.~K.}\ \bibnamefont {{Berggren}}},
  \bibinfo {author} {\bibfnamefont {D.}~\bibnamefont {{Braga}}}, \bibinfo
  {author} {\bibfnamefont {S.~A.}\ \bibnamefont {{Charlebois}}}, \bibinfo
  {author} {\bibfnamefont {C.}~\bibnamefont {{Chang}}}, \bibinfo {author}
  {\bibfnamefont {A.}~\bibnamefont {{Dragone}}}, \bibinfo {author}
  {\bibfnamefont {A.}~\bibnamefont {{Drlica-Wagner}}}, \bibinfo {author}
  {\bibfnamefont {C.~O.}\ \bibnamefont {{Escobar}}}, \bibinfo {author}
  {\bibfnamefont {J.}~\bibnamefont {{Estrada}}}, \bibinfo {author}
  {\bibfnamefont {F.}~\bibnamefont {{Fahim}}}, \bibinfo {author} {\bibfnamefont
  {M.}~\bibnamefont {{Febbraro}}}, \bibinfo {author} {\bibfnamefont
  {G.}~\bibnamefont {{Fernandez Moroni}}}, \bibinfo {author} {\bibfnamefont
  {S.}~\bibnamefont {{Holland}}}, \bibinfo {author} {\bibfnamefont
  {T.}~\bibnamefont {{Hossbach}}}, \bibinfo {author} {\bibfnamefont
  {S.}~\bibnamefont {{Koppell}}}, \bibinfo {author} {\bibfnamefont
  {C.}~\bibnamefont {{Leitz}}}, \bibinfo {author} {\bibfnamefont
  {A.}~\bibnamefont {{Magnoni}}}, \bibinfo {author} {\bibfnamefont {B.~A.}\
  \bibnamefont {{Mazin}}}, \bibinfo {author} {\bibfnamefont {J.-F.}\
  \bibnamefont {{Pratte}}}, \bibinfo {author} {\bibfnamefont {B.}~\bibnamefont
  {{Rauscher}}}, \bibinfo {author} {\bibfnamefont {D.}~\bibnamefont
  {{Rodrigues}}}, \bibinfo {author} {\bibfnamefont {L.}~\bibnamefont {{Shen}}},
  \bibinfo {author} {\bibfnamefont {M.}~\bibnamefont {{Sofo-Haro}}}, \bibinfo
  {author} {\bibfnamefont {J.}~\bibnamefont {{Tiffenberg}}}, \bibinfo {author}
  {\bibfnamefont {J.}~\bibnamefont {{Turner}}}, \bibinfo {author}
  {\bibfnamefont {L.}~\bibnamefont {{Rota}}}, \bibinfo {author} {\bibfnamefont
  {C.~J.}\ \bibnamefont {{Kenney}}}, \bibinfo {author} {\bibfnamefont
  {F.}~\bibnamefont {{Vachon}}}, \ and\ \bibinfo {author} {\bibfnamefont
  {G.}~\bibnamefont {{Wang}}},\ }\href {\doibase 10.48550/arXiv.2203.12542}
  {\bibfield  {journal} {\bibinfo  {journal} {arXiv e-prints}\ ,\ \bibinfo
  {eid} {arXiv:2203.12542}} (\bibinfo {year} {2022})},\ \Eprint
  {http://arxiv.org/abs/2203.12542} {arXiv:2203.12542 [physics.ins-det]}
  \BibitemShut {NoStop}%
\bibitem [{\citenamefont {Yuan}\ \emph {et~al.}(2010)\citenamefont {Yuan},
  \citenamefont {Bao}, \citenamefont {Lu}, \citenamefont {Zhang}, \citenamefont
  {Peng},\ and\ \citenamefont {Pan}}]{YUAN20101}%
  \BibitemOpen
  \bibfield  {author} {\bibinfo {author} {\bibfnamefont {Z.-S.}\ \bibnamefont
  {Yuan}}, \bibinfo {author} {\bibfnamefont {X.-H.}\ \bibnamefont {Bao}},
  \bibinfo {author} {\bibfnamefont {C.-Y.}\ \bibnamefont {Lu}}, \bibinfo
  {author} {\bibfnamefont {J.}~\bibnamefont {Zhang}}, \bibinfo {author}
  {\bibfnamefont {C.-Z.}\ \bibnamefont {Peng}}, \ and\ \bibinfo {author}
  {\bibfnamefont {J.-W.}\ \bibnamefont {Pan}},\ }\href {\doibase
  https://doi.org/10.1016/j.physrep.2010.07.004} {\bibfield  {journal}
  {\bibinfo  {journal} {Physics Reports}\ }\textbf {\bibinfo {volume} {497}},\
  \bibinfo {pages} {1} (\bibinfo {year} {2010})}\BibitemShut {NoStop}%
\bibitem [{\citenamefont {{Malygin}}\ \emph {et~al.}(1985)\citenamefont
  {{Malygin}}, \citenamefont {{Penin}},\ and\ \citenamefont
  {{Sergienko}}}]{malygin_1985}%
  \BibitemOpen
  \bibfield  {author} {\bibinfo {author} {\bibfnamefont {A.~A.}\ \bibnamefont
  {{Malygin}}}, \bibinfo {author} {\bibfnamefont {A.~N.}\ \bibnamefont
  {{Penin}}}, \ and\ \bibinfo {author} {\bibfnamefont {A.~V.}\ \bibnamefont
  {{Sergienko}}},\ }\href@noop {} {\bibfield  {journal} {\bibinfo  {journal}
  {Soviet Physics Doklady}\ }\textbf {\bibinfo {volume} {20}},\ \bibinfo
  {pages} {227} (\bibinfo {year} {1985})}\BibitemShut {NoStop}%
\bibitem [{\citenamefont {Padgett}\ and\ \citenamefont
  {Boyd}(2017)}]{Padgett_2017}%
  \BibitemOpen
  \bibfield  {author} {\bibinfo {author} {\bibfnamefont {M.~J.}\ \bibnamefont
  {Padgett}}\ and\ \bibinfo {author} {\bibfnamefont {R.~W.}\ \bibnamefont
  {Boyd}},\ }\href@noop {} {\bibfield  {journal} {\bibinfo  {journal}
  {Philosophical Transactions of the Royal Society A: Mathematical, Physical
  and Engineering Sciences}\ }\textbf {\bibinfo {volume} {375}},\ \bibinfo
  {pages} {20160233} (\bibinfo {year} {2017})}\BibitemShut {NoStop}%
\bibitem [{\citenamefont {Pittman}\ \emph {et~al.}(1995)\citenamefont
  {Pittman}, \citenamefont {Shih}, \citenamefont {Strekalov},\ and\
  \citenamefont {Sergienko}}]{Pittman_1995}%
  \BibitemOpen
  \bibfield  {author} {\bibinfo {author} {\bibfnamefont {T.~B.}\ \bibnamefont
  {Pittman}}, \bibinfo {author} {\bibfnamefont {Y.~H.}\ \bibnamefont {Shih}},
  \bibinfo {author} {\bibfnamefont {D.~V.}\ \bibnamefont {Strekalov}}, \ and\
  \bibinfo {author} {\bibfnamefont {A.~V.}\ \bibnamefont {Sergienko}},\ }\href
  {\doibase 10.1103/PhysRevA.52.R3429} {\bibfield  {journal} {\bibinfo
  {journal} {Phys. Rev. A}\ }\textbf {\bibinfo {volume} {52}},\ \bibinfo
  {pages} {R3429} (\bibinfo {year} {1995})}\BibitemShut {NoStop}%
\bibitem [{\citenamefont {Defienne}\ \emph {et~al.}(2018)\citenamefont
  {Defienne}, \citenamefont {Reichert},\ and\ \citenamefont
  {Fleischer}}]{defienne_2010}%
  \BibitemOpen
  \bibfield  {author} {\bibinfo {author} {\bibfnamefont {H.}~\bibnamefont
  {Defienne}}, \bibinfo {author} {\bibfnamefont {M.}~\bibnamefont {Reichert}},
  \ and\ \bibinfo {author} {\bibfnamefont {J.~W.}\ \bibnamefont {Fleischer}},\
  }\href {\doibase 10.1103/PhysRevLett.120.203604} {\bibfield  {journal}
  {\bibinfo  {journal} {Phys. Rev. Lett.}\ }\textbf {\bibinfo {volume} {120}},\
  \bibinfo {pages} {203604} (\bibinfo {year} {2018})}\BibitemShut {NoStop}%
\bibitem [{\citenamefont {Bestvater}\ \emph {et~al.}(2010)\citenamefont
  {Bestvater}, \citenamefont {Seghiri}, \citenamefont {Kang}, \citenamefont
  {Gröner}, \citenamefont {Lee}, \citenamefont {Im},\ and\ \citenamefont
  {Wachsmuth}}]{Bestvater_2010}%
  \BibitemOpen
  \bibfield  {author} {\bibinfo {author} {\bibfnamefont {F.}~\bibnamefont
  {Bestvater}}, \bibinfo {author} {\bibfnamefont {Z.}~\bibnamefont {Seghiri}},
  \bibinfo {author} {\bibfnamefont {M.}~\bibnamefont {Kang}}, \bibinfo {author}
  {\bibfnamefont {N.}~\bibnamefont {Gröner}}, \bibinfo {author} {\bibfnamefont
  {J.-Y.}\ \bibnamefont {Lee}}, \bibinfo {author} {\bibfnamefont {K.-B.}\
  \bibnamefont {Im}}, \ and\ \bibinfo {author} {\bibfnamefont {M.}~\bibnamefont
  {Wachsmuth}},\ }\href {\doibase 10.1364/OE.18.023818} {\bibfield  {journal}
  {\bibinfo  {journal} {Optics express}\ }\textbf {\bibinfo {volume} {18}},\
  \bibinfo {pages} {23818} (\bibinfo {year} {2010})}\BibitemShut {NoStop}%
\bibitem [{\citenamefont {Orieux}\ \emph {et~al.}(2017)\citenamefont {Orieux},
  \citenamefont {Versteegh}, \citenamefont {JÃ¶ns},\ and\ \citenamefont
  {Ducci}}]{Orieux_2017}%
  \BibitemOpen
  \bibfield  {author} {\bibinfo {author} {\bibfnamefont {A.}~\bibnamefont
  {Orieux}}, \bibinfo {author} {\bibfnamefont {M.~A.~M.}\ \bibnamefont
  {Versteegh}}, \bibinfo {author} {\bibfnamefont {K.~D.}\ \bibnamefont
  {JÃ¶ns}}, \ and\ \bibinfo {author} {\bibfnamefont {S.}~\bibnamefont
  {Ducci}},\ }\href {\doibase 10.1088/1361-6633/aa6955} {\bibfield  {journal}
  {\bibinfo  {journal} {Reports on Progress in Physics}\ }\textbf {\bibinfo
  {volume} {80}},\ \bibinfo {pages} {076001} (\bibinfo {year}
  {2017})}\BibitemShut {NoStop}%
\bibitem [{\citenamefont {Aguilar-Arevalo}\ \emph {et~al.}(2019)\citenamefont
  {Aguilar-Arevalo}, \citenamefont {Amidei}, \citenamefont {Baxter},
  \citenamefont {Cancelo}, \citenamefont {Cervantes~Vergara}, \citenamefont
  {Chavarria}, \citenamefont {Darragh-Ford}, \citenamefont {de~Mello~Neto},
  \citenamefont {D'Olivo}, \citenamefont {Estrada}, \citenamefont {Ga\"{\i}or},
  \citenamefont {Guardincerri}, \citenamefont {Hossbach}, \citenamefont
  {Kilminster}, \citenamefont {Lawson}, \citenamefont {Lee}, \citenamefont
  {Letessier-Selvon}, \citenamefont {Matalon}, \citenamefont {Mello},
  \citenamefont {Mitra}, \citenamefont {Molina}, \citenamefont {Paul},
  \citenamefont {Piers}, \citenamefont {Privitera}, \citenamefont {Ramanathan},
  \citenamefont {Da~Rocha}, \citenamefont {Sarkis}, \citenamefont {Settimo},
  \citenamefont {Smida}, \citenamefont {Thomas}, \citenamefont {Tiffenberg},
  \citenamefont {Torres~Machado}, \citenamefont {Vilar},\ and\ \citenamefont
  {Virto}}]{DAMIC2019}%
  \BibitemOpen
  \bibfield  {author} {\bibinfo {author} {\bibfnamefont {A.}~\bibnamefont
  {Aguilar-Arevalo}}, \bibinfo {author} {\bibfnamefont {D.}~\bibnamefont
  {Amidei}}, \bibinfo {author} {\bibfnamefont {D.}~\bibnamefont {Baxter}},
  \bibinfo {author} {\bibfnamefont {G.}~\bibnamefont {Cancelo}}, \bibinfo
  {author} {\bibfnamefont {B.~A.}\ \bibnamefont {Cervantes~Vergara}}, \bibinfo
  {author} {\bibfnamefont {A.~E.}\ \bibnamefont {Chavarria}}, \bibinfo {author}
  {\bibfnamefont {E.}~\bibnamefont {Darragh-Ford}}, \bibinfo {author}
  {\bibfnamefont {J.~R.~T.}\ \bibnamefont {de~Mello~Neto}}, \bibinfo {author}
  {\bibfnamefont {J.~C.}\ \bibnamefont {D'Olivo}}, \bibinfo {author}
  {\bibfnamefont {J.}~\bibnamefont {Estrada}}, \bibinfo {author} {\bibfnamefont
  {R.}~\bibnamefont {Ga\"{\i}or}}, \bibinfo {author} {\bibfnamefont
  {Y.}~\bibnamefont {Guardincerri}}, \bibinfo {author} {\bibfnamefont {T.~W.}\
  \bibnamefont {Hossbach}}, \bibinfo {author} {\bibfnamefont {B.}~\bibnamefont
  {Kilminster}}, \bibinfo {author} {\bibfnamefont {I.}~\bibnamefont {Lawson}},
  \bibinfo {author} {\bibfnamefont {S.~J.}\ \bibnamefont {Lee}}, \bibinfo
  {author} {\bibfnamefont {A.}~\bibnamefont {Letessier-Selvon}}, \bibinfo
  {author} {\bibfnamefont {A.}~\bibnamefont {Matalon}}, \bibinfo {author}
  {\bibfnamefont {V.~B.~B.}\ \bibnamefont {Mello}}, \bibinfo {author}
  {\bibfnamefont {P.}~\bibnamefont {Mitra}}, \bibinfo {author} {\bibfnamefont
  {J.}~\bibnamefont {Molina}}, \bibinfo {author} {\bibfnamefont
  {S.}~\bibnamefont {Paul}}, \bibinfo {author} {\bibfnamefont {A.}~\bibnamefont
  {Piers}}, \bibinfo {author} {\bibfnamefont {P.}~\bibnamefont {Privitera}},
  \bibinfo {author} {\bibfnamefont {K.}~\bibnamefont {Ramanathan}}, \bibinfo
  {author} {\bibfnamefont {J.}~\bibnamefont {Da~Rocha}}, \bibinfo {author}
  {\bibfnamefont {Y.}~\bibnamefont {Sarkis}}, \bibinfo {author} {\bibfnamefont
  {M.}~\bibnamefont {Settimo}}, \bibinfo {author} {\bibfnamefont
  {R.}~\bibnamefont {Smida}}, \bibinfo {author} {\bibfnamefont
  {R.}~\bibnamefont {Thomas}}, \bibinfo {author} {\bibfnamefont
  {J.}~\bibnamefont {Tiffenberg}}, \bibinfo {author} {\bibfnamefont
  {D.}~\bibnamefont {Torres~Machado}}, \bibinfo {author} {\bibfnamefont
  {R.}~\bibnamefont {Vilar}}, \ and\ \bibinfo {author} {\bibfnamefont {A.~L.}\
  \bibnamefont {Virto}} (\bibinfo {collaboration} {DAMIC Collaboration}),\
  }\href {\doibase 10.1103/PhysRevLett.123.181802} {\bibfield  {journal}
  {\bibinfo  {journal} {Phys. Rev. Lett.}\ }\textbf {\bibinfo {volume} {123}},\
  \bibinfo {pages} {181802} (\bibinfo {year} {2019})}\BibitemShut {NoStop}%
\bibitem [{\citenamefont {Arnquist}\ \emph {et~al.}(2023)\citenamefont
  {Arnquist}, \citenamefont {Avalos}, \citenamefont {Baxter}, \citenamefont
  {Bertou}, \citenamefont {Castell\'o-Mor}, \citenamefont {Chavarria},
  \citenamefont {Cuevas-Zepeda}, \citenamefont {Guti\'errez}, \citenamefont
  {Duarte-Campderros}, \citenamefont {Dastgheibi-Fard}, \citenamefont
  {Deligny}, \citenamefont {De~Dominicis}, \citenamefont {Estrada},
  \citenamefont {Gadola}, \citenamefont {Ga\"{\i}or}, \citenamefont {Hossbach},
  \citenamefont {Iddir}, \citenamefont {Khalil}, \citenamefont {Kilminster},
  \citenamefont {Lantero-Barreda}, \citenamefont {Lawson}, \citenamefont {Lee},
  \citenamefont {Letessier-Selvon}, \citenamefont {Loaiza}, \citenamefont
  {Lopez-Virto}, \citenamefont {Matalon}, \citenamefont {Munagavalasa},
  \citenamefont {McGuire}, \citenamefont {Mitra}, \citenamefont {Norcini},
  \citenamefont {Papadopoulos}, \citenamefont {Paul}, \citenamefont {Piers},
  \citenamefont {Privitera}, \citenamefont {Ramanathan}, \citenamefont
  {Robmann}, \citenamefont {Settimo}, \citenamefont {Smida}, \citenamefont
  {Thomas}, \citenamefont {Traina}, \citenamefont {Vila}, \citenamefont
  {Vilar}, \citenamefont {Warot}, \citenamefont {Yajur},\ and\ \citenamefont
  {Zopounidis}}]{damicM_2023}%
  \BibitemOpen
  \bibfield  {author} {\bibinfo {author} {\bibfnamefont {I.}~\bibnamefont
  {Arnquist}}, \bibinfo {author} {\bibfnamefont {N.}~\bibnamefont {Avalos}},
  \bibinfo {author} {\bibfnamefont {D.}~\bibnamefont {Baxter}}, \bibinfo
  {author} {\bibfnamefont {X.}~\bibnamefont {Bertou}}, \bibinfo {author}
  {\bibfnamefont {N.}~\bibnamefont {Castell\'o-Mor}}, \bibinfo {author}
  {\bibfnamefont {A.~E.}\ \bibnamefont {Chavarria}}, \bibinfo {author}
  {\bibfnamefont {J.}~\bibnamefont {Cuevas-Zepeda}}, \bibinfo {author}
  {\bibfnamefont {J.~C.}\ \bibnamefont {Guti\'errez}}, \bibinfo {author}
  {\bibfnamefont {J.}~\bibnamefont {Duarte-Campderros}}, \bibinfo {author}
  {\bibfnamefont {A.}~\bibnamefont {Dastgheibi-Fard}}, \bibinfo {author}
  {\bibfnamefont {O.}~\bibnamefont {Deligny}}, \bibinfo {author} {\bibfnamefont
  {C.}~\bibnamefont {De~Dominicis}}, \bibinfo {author} {\bibfnamefont
  {E.}~\bibnamefont {Estrada}}, \bibinfo {author} {\bibfnamefont
  {N.}~\bibnamefont {Gadola}}, \bibinfo {author} {\bibfnamefont
  {R.}~\bibnamefont {Ga\"{\i}or}}, \bibinfo {author} {\bibfnamefont
  {T.}~\bibnamefont {Hossbach}}, \bibinfo {author} {\bibfnamefont
  {L.}~\bibnamefont {Iddir}}, \bibinfo {author} {\bibfnamefont
  {L.}~\bibnamefont {Khalil}}, \bibinfo {author} {\bibfnamefont
  {B.}~\bibnamefont {Kilminster}}, \bibinfo {author} {\bibfnamefont
  {A.}~\bibnamefont {Lantero-Barreda}}, \bibinfo {author} {\bibfnamefont
  {I.}~\bibnamefont {Lawson}}, \bibinfo {author} {\bibfnamefont
  {S.}~\bibnamefont {Lee}}, \bibinfo {author} {\bibfnamefont {A.}~\bibnamefont
  {Letessier-Selvon}}, \bibinfo {author} {\bibfnamefont {P.}~\bibnamefont
  {Loaiza}}, \bibinfo {author} {\bibfnamefont {A.}~\bibnamefont {Lopez-Virto}},
  \bibinfo {author} {\bibfnamefont {A.}~\bibnamefont {Matalon}}, \bibinfo
  {author} {\bibfnamefont {S.}~\bibnamefont {Munagavalasa}}, \bibinfo {author}
  {\bibfnamefont {K.~J.}\ \bibnamefont {McGuire}}, \bibinfo {author}
  {\bibfnamefont {P.}~\bibnamefont {Mitra}}, \bibinfo {author} {\bibfnamefont
  {D.}~\bibnamefont {Norcini}}, \bibinfo {author} {\bibfnamefont
  {G.}~\bibnamefont {Papadopoulos}}, \bibinfo {author} {\bibfnamefont
  {S.}~\bibnamefont {Paul}}, \bibinfo {author} {\bibfnamefont {A.}~\bibnamefont
  {Piers}}, \bibinfo {author} {\bibfnamefont {P.}~\bibnamefont {Privitera}},
  \bibinfo {author} {\bibfnamefont {K.}~\bibnamefont {Ramanathan}}, \bibinfo
  {author} {\bibfnamefont {P.}~\bibnamefont {Robmann}}, \bibinfo {author}
  {\bibfnamefont {M.}~\bibnamefont {Settimo}}, \bibinfo {author} {\bibfnamefont
  {R.}~\bibnamefont {Smida}}, \bibinfo {author} {\bibfnamefont
  {R.}~\bibnamefont {Thomas}}, \bibinfo {author} {\bibfnamefont
  {M.}~\bibnamefont {Traina}}, \bibinfo {author} {\bibfnamefont
  {I.}~\bibnamefont {Vila}}, \bibinfo {author} {\bibfnamefont {R.}~\bibnamefont
  {Vilar}}, \bibinfo {author} {\bibfnamefont {G.}~\bibnamefont {Warot}},
  \bibinfo {author} {\bibfnamefont {R.}~\bibnamefont {Yajur}}, \ and\ \bibinfo
  {author} {\bibfnamefont {J.-P.}\ \bibnamefont {Zopounidis}} (\bibinfo
  {collaboration} {DAMIC-M Collaboration}),\ }\href {\doibase
  10.1103/PhysRevLett.130.171003} {\bibfield  {journal} {\bibinfo  {journal}
  {Phys. Rev. Lett.}\ }\textbf {\bibinfo {volume} {130}},\ \bibinfo {pages}
  {171003} (\bibinfo {year} {2023})}\BibitemShut {NoStop}%
\bibitem [{\citenamefont {{Aguilar-Arevalo}}\ \emph {et~al.}(2022)\citenamefont
  {{Aguilar-Arevalo}}, \citenamefont {{Alcalde Bessia}}, \citenamefont
  {{Avalos}}, \citenamefont {{Baxter}}, \citenamefont {{Bertou}}, \citenamefont
  {{Bonifazi}}, \citenamefont {{Botti}}, \citenamefont {{Cababie}},
  \citenamefont {{Cancelo}}, \citenamefont {{Cervantes-Vergara}}, \citenamefont
  {{Castello-Mor}}, \citenamefont {{Chavarria}}, \citenamefont {{Chavez}},
  \citenamefont {{Chierchie}}, \citenamefont {{De Egea}}, \citenamefont
  {{D`Olivo}}, \citenamefont {{Dreyer}}, \citenamefont {{Drlica-Wagner}},
  \citenamefont {{Essig}}, \citenamefont {{Estrada}}, \citenamefont
  {{Estrada}}, \citenamefont {{Etzion}}, \citenamefont {{Fernandez-Moroni}},
  \citenamefont {{Fernandez-Serra}}, \citenamefont {{Holland}}, \citenamefont
  {{Lantero Barreda}}, \citenamefont {{Lathrop}}, \citenamefont {{Lipovetzky}},
  \citenamefont {{Loer}}, \citenamefont {{Marrufo Villalpando}}, \citenamefont
  {{Molina}}, \citenamefont {{Perez}}, \citenamefont {{Privitera}},
  \citenamefont {{Rodrigues}}, \citenamefont {{Saldanha}}, \citenamefont
  {{Santa Cruz}}, \citenamefont {{Singal}}, \citenamefont {{Saffold}},
  \citenamefont {{Stefanazzi}}, \citenamefont {{Sofo-Haro}}, \citenamefont
  {{Tiffenberg}}, \citenamefont {{Torres}}, \citenamefont {{Uemura}},\ and\
  \citenamefont {{Vilar}}}]{oscura2022}%
  \BibitemOpen
  \bibfield  {author} {\bibinfo {author} {\bibfnamefont {A.}~\bibnamefont
  {{Aguilar-Arevalo}}}, \bibinfo {author} {\bibfnamefont {F.}~\bibnamefont
  {{Alcalde Bessia}}}, \bibinfo {author} {\bibfnamefont {N.}~\bibnamefont
  {{Avalos}}}, \bibinfo {author} {\bibfnamefont {D.}~\bibnamefont {{Baxter}}},
  \bibinfo {author} {\bibfnamefont {X.}~\bibnamefont {{Bertou}}}, \bibinfo
  {author} {\bibfnamefont {C.}~\bibnamefont {{Bonifazi}}}, \bibinfo {author}
  {\bibfnamefont {A.}~\bibnamefont {{Botti}}}, \bibinfo {author} {\bibfnamefont
  {M.}~\bibnamefont {{Cababie}}}, \bibinfo {author} {\bibfnamefont
  {G.}~\bibnamefont {{Cancelo}}}, \bibinfo {author} {\bibfnamefont {B.~A.}\
  \bibnamefont {{Cervantes-Vergara}}}, \bibinfo {author} {\bibfnamefont
  {N.}~\bibnamefont {{Castello-Mor}}}, \bibinfo {author} {\bibfnamefont
  {A.}~\bibnamefont {{Chavarria}}}, \bibinfo {author} {\bibfnamefont {C.~R.}\
  \bibnamefont {{Chavez}}}, \bibinfo {author} {\bibfnamefont {F.}~\bibnamefont
  {{Chierchie}}}, \bibinfo {author} {\bibfnamefont {J.~M.}\ \bibnamefont {{De
  Egea}}}, \bibinfo {author} {\bibfnamefont {J.~C.}\ \bibnamefont {{D`Olivo}}},
  \bibinfo {author} {\bibfnamefont {C.~E.}\ \bibnamefont {{Dreyer}}}, \bibinfo
  {author} {\bibfnamefont {A.}~\bibnamefont {{Drlica-Wagner}}}, \bibinfo
  {author} {\bibfnamefont {R.}~\bibnamefont {{Essig}}}, \bibinfo {author}
  {\bibfnamefont {J.}~\bibnamefont {{Estrada}}}, \bibinfo {author}
  {\bibfnamefont {E.}~\bibnamefont {{Estrada}}}, \bibinfo {author}
  {\bibfnamefont {E.}~\bibnamefont {{Etzion}}}, \bibinfo {author}
  {\bibfnamefont {G.}~\bibnamefont {{Fernandez-Moroni}}}, \bibinfo {author}
  {\bibfnamefont {M.}~\bibnamefont {{Fernandez-Serra}}}, \bibinfo {author}
  {\bibfnamefont {S.}~\bibnamefont {{Holland}}}, \bibinfo {author}
  {\bibfnamefont {A.}~\bibnamefont {{Lantero Barreda}}}, \bibinfo {author}
  {\bibfnamefont {A.}~\bibnamefont {{Lathrop}}}, \bibinfo {author}
  {\bibfnamefont {J.}~\bibnamefont {{Lipovetzky}}}, \bibinfo {author}
  {\bibfnamefont {B.}~\bibnamefont {{Loer}}}, \bibinfo {author} {\bibfnamefont
  {E.}~\bibnamefont {{Marrufo Villalpando}}}, \bibinfo {author} {\bibfnamefont
  {J.}~\bibnamefont {{Molina}}}, \bibinfo {author} {\bibfnamefont
  {S.}~\bibnamefont {{Perez}}}, \bibinfo {author} {\bibfnamefont
  {P.}~\bibnamefont {{Privitera}}}, \bibinfo {author} {\bibfnamefont
  {D.}~\bibnamefont {{Rodrigues}}}, \bibinfo {author} {\bibfnamefont
  {R.}~\bibnamefont {{Saldanha}}}, \bibinfo {author} {\bibfnamefont
  {D.}~\bibnamefont {{Santa Cruz}}}, \bibinfo {author} {\bibfnamefont
  {A.}~\bibnamefont {{Singal}}}, \bibinfo {author} {\bibfnamefont
  {N.}~\bibnamefont {{Saffold}}}, \bibinfo {author} {\bibfnamefont
  {L.}~\bibnamefont {{Stefanazzi}}}, \bibinfo {author} {\bibfnamefont
  {M.}~\bibnamefont {{Sofo-Haro}}}, \bibinfo {author} {\bibfnamefont
  {J.}~\bibnamefont {{Tiffenberg}}}, \bibinfo {author} {\bibfnamefont
  {C.}~\bibnamefont {{Torres}}}, \bibinfo {author} {\bibfnamefont
  {S.}~\bibnamefont {{Uemura}}}, \ and\ \bibinfo {author} {\bibfnamefont
  {R.}~\bibnamefont {{Vilar}}},\ }\href {\doibase 10.48550/arXiv.2202.10518}
  {\bibfield  {journal} {\bibinfo  {journal} {arXiv e-prints}\ ,\ \bibinfo
  {eid} {arXiv:2202.10518}} (\bibinfo {year} {2022})},\ \Eprint
  {http://arxiv.org/abs/2202.10518} {arXiv:2202.10518 [astro-ph.IM]}
  \BibitemShut {NoStop}%
\bibitem [{\citenamefont {Aguilar-Arevalo}\ \emph {et~al.}(2022)\citenamefont
  {Aguilar-Arevalo}, \citenamefont {Bernal}, \citenamefont {Bertou},
  \citenamefont {Bonifazi}, \citenamefont {Cancelo}, \citenamefont
  {de~Carvalho}, \citenamefont {Cervantes-Vergara}, \citenamefont {Chavez},
  \citenamefont {Corr{\^e}a}, \citenamefont {D'Olivo}, \citenamefont {dos
  Anjos}, \citenamefont {Estrada}, \citenamefont {Fernandes~Neto},
  \citenamefont {Fernandez~Moroni}, \citenamefont {Foguel}, \citenamefont
  {Ford}, \citenamefont {Gasanego~Barbuscio}, \citenamefont {Gonzalez~Cuevas},
  \citenamefont {Hernandez}, \citenamefont {Izraelevitch}, \citenamefont
  {Kilminster}, \citenamefont {Kuk}, \citenamefont {Lima}, \citenamefont
  {Makler}, \citenamefont {Martinez~Montero}, \citenamefont {Mendes},
  \citenamefont {Molina}, \citenamefont {Mota}, \citenamefont {Nasteva},
  \citenamefont {Paolini}, \citenamefont {Rodrigues}, \citenamefont {Sarkis},
  \citenamefont {Sofo~Haro}, \citenamefont {Stalder}, \citenamefont
  {Tiffenberg},\ and\ \citenamefont {collaboration}}]{connie_2022}%
  \BibitemOpen
  \bibfield  {author} {\bibinfo {author} {\bibfnamefont {A.}~\bibnamefont
  {Aguilar-Arevalo}}, \bibinfo {author} {\bibfnamefont {J.}~\bibnamefont
  {Bernal}}, \bibinfo {author} {\bibfnamefont {X.}~\bibnamefont {Bertou}},
  \bibinfo {author} {\bibfnamefont {C.}~\bibnamefont {Bonifazi}}, \bibinfo
  {author} {\bibfnamefont {G.}~\bibnamefont {Cancelo}}, \bibinfo {author}
  {\bibfnamefont {V.~G. P.~B.}\ \bibnamefont {de~Carvalho}}, \bibinfo {author}
  {\bibfnamefont {B.~A.}\ \bibnamefont {Cervantes-Vergara}}, \bibinfo {author}
  {\bibfnamefont {C.}~\bibnamefont {Chavez}}, \bibinfo {author} {\bibfnamefont
  {G.~C.}\ \bibnamefont {Corr{\^e}a}}, \bibinfo {author} {\bibfnamefont
  {J.~C.}\ \bibnamefont {D'Olivo}}, \bibinfo {author} {\bibfnamefont {J.~C.}\
  \bibnamefont {dos Anjos}}, \bibinfo {author} {\bibfnamefont {J.}~\bibnamefont
  {Estrada}}, \bibinfo {author} {\bibfnamefont {A.~R.}\ \bibnamefont
  {Fernandes~Neto}}, \bibinfo {author} {\bibfnamefont {G.}~\bibnamefont
  {Fernandez~Moroni}}, \bibinfo {author} {\bibfnamefont {A.}~\bibnamefont
  {Foguel}}, \bibinfo {author} {\bibfnamefont {R.}~\bibnamefont {Ford}},
  \bibinfo {author} {\bibfnamefont {J.}~\bibnamefont {Gasanego~Barbuscio}},
  \bibinfo {author} {\bibfnamefont {J.}~\bibnamefont {Gonzalez~Cuevas}},
  \bibinfo {author} {\bibfnamefont {S.}~\bibnamefont {Hernandez}}, \bibinfo
  {author} {\bibfnamefont {F.}~\bibnamefont {Izraelevitch}}, \bibinfo {author}
  {\bibfnamefont {B.}~\bibnamefont {Kilminster}}, \bibinfo {author}
  {\bibfnamefont {K.}~\bibnamefont {Kuk}}, \bibinfo {author} {\bibfnamefont
  {H.~P.}\ \bibnamefont {Lima}}, \bibinfo {author} {\bibfnamefont
  {M.}~\bibnamefont {Makler}}, \bibinfo {author} {\bibfnamefont
  {M.}~\bibnamefont {Martinez~Montero}}, \bibinfo {author} {\bibfnamefont
  {L.~H.}\ \bibnamefont {Mendes}}, \bibinfo {author} {\bibfnamefont
  {J.}~\bibnamefont {Molina}}, \bibinfo {author} {\bibfnamefont
  {P.}~\bibnamefont {Mota}}, \bibinfo {author} {\bibfnamefont {I.}~\bibnamefont
  {Nasteva}}, \bibinfo {author} {\bibfnamefont {E.}~\bibnamefont {Paolini}},
  \bibinfo {author} {\bibfnamefont {D.}~\bibnamefont {Rodrigues}}, \bibinfo
  {author} {\bibfnamefont {Y.}~\bibnamefont {Sarkis}}, \bibinfo {author}
  {\bibfnamefont {M.}~\bibnamefont {Sofo~Haro}}, \bibinfo {author}
  {\bibfnamefont {D.}~\bibnamefont {Stalder}}, \bibinfo {author} {\bibfnamefont
  {J.}~\bibnamefont {Tiffenberg}}, \ and\ \bibinfo {author} {\bibfnamefont
  {T.~C.}\ \bibnamefont {collaboration}},\ }\href {\doibase
  10.1007/JHEP05(2022)017} {\bibfield  {journal} {\bibinfo  {journal} {Journal
  of High Energy Physics}\ }\textbf {\bibinfo {volume} {2022}},\ \bibinfo
  {pages} {17} (\bibinfo {year} {2022})}\BibitemShut {NoStop}%
\bibitem [{\citenamefont {Moroni}\ \emph {et~al.}(2022)\citenamefont {Moroni},
  \citenamefont {Chierchie}, \citenamefont {Tiffenberg}, \citenamefont {Botti},
  \citenamefont {Cababie}, \citenamefont {Cancelo}, \citenamefont {Depaoli},
  \citenamefont {Estrada}, \citenamefont {Holland}, \citenamefont {Rodrigues},
  \citenamefont {Sidelnik}, \citenamefont {Haro}, \citenamefont {Stefanazzi},\
  and\ \citenamefont {Uemura}}]{skipper_above_ground_2022}%
  \BibitemOpen
  \bibfield  {author} {\bibinfo {author} {\bibfnamefont {G.~F.}\ \bibnamefont
  {Moroni}}, \bibinfo {author} {\bibfnamefont {F.}~\bibnamefont {Chierchie}},
  \bibinfo {author} {\bibfnamefont {J.}~\bibnamefont {Tiffenberg}}, \bibinfo
  {author} {\bibfnamefont {A.}~\bibnamefont {Botti}}, \bibinfo {author}
  {\bibfnamefont {M.}~\bibnamefont {Cababie}}, \bibinfo {author} {\bibfnamefont
  {G.}~\bibnamefont {Cancelo}}, \bibinfo {author} {\bibfnamefont {E.~L.}\
  \bibnamefont {Depaoli}}, \bibinfo {author} {\bibfnamefont {J.}~\bibnamefont
  {Estrada}}, \bibinfo {author} {\bibfnamefont {S.~E.}\ \bibnamefont
  {Holland}}, \bibinfo {author} {\bibfnamefont {D.}~\bibnamefont {Rodrigues}},
  \bibinfo {author} {\bibfnamefont {I.}~\bibnamefont {Sidelnik}}, \bibinfo
  {author} {\bibfnamefont {M.~S.}\ \bibnamefont {Haro}}, \bibinfo {author}
  {\bibfnamefont {L.}~\bibnamefont {Stefanazzi}}, \ and\ \bibinfo {author}
  {\bibfnamefont {S.}~\bibnamefont {Uemura}},\ }\href {\doibase
  10.1103/PhysRevApplied.17.044050} {\bibfield  {journal} {\bibinfo  {journal}
  {Phys. Rev. Appl.}\ }\textbf {\bibinfo {volume} {17}},\ \bibinfo {pages}
  {044050} (\bibinfo {year} {2022})}\BibitemShut {NoStop}%
\bibitem [{\citenamefont {IARPA}()}]{grail}%
  \BibitemOpen
  \bibfield  {author} {\bibinfo {author} {\bibnamefont {IARPA}},\ }\href
  {https://www.iarpa.gov/research-programs/grail} {\enquote {\bibinfo {title}
  {G{RAIL}: Gaseous radioisotope analysis in situ laboratory},}\ }\BibitemShut
  {NoStop}%
\end{thebibliography}%
\bibliographystyle{apsrev4-1}

\end{document}